\documentclass[twocolumn,twocolappendix]{aastex63}

\usepackage{newtxtext,newtxmath}

\usepackage[T1]{fontenc}
\usepackage{ae,aecompl}
\usepackage{comment}

\usepackage{graphicx,tabularx}	
\usepackage{amsmath}	
\usepackage{amssymb}	
\usepackage{booktabs}
\usepackage[inline,shortlabels]{enumitem}
\setlist{itemjoin* = { and\enspace}}
\usepackage{color}
\usepackage{multirow}
\usepackage{natbib}
\usepackage[normalem]{ulem}
\usepackage{cancel}

\defcitealias{Gillet2019}{G19}
\defcitealias{alsing2019fast}{A19}
\received{07-May-2021}
\revised{08-Dec-2021}
\accepted{19-Dec-2021}
\submitjournal{ApJ}

\shorttitle{Reionization parameter inference}
\shortauthors{X.~Zhao et al.}
\graphicspath{{./}}

\begin{document}
\title{Simulation-Based Inference of Reionization Parameters From 3D Tomographic 21 cm Lightcone Images} 

\correspondingauthor{Yi Mao}\email{ymao@tsinghua.edu.cn}

\author[0000-0002-8328-1447]{Xiaosheng Zhao}
\affiliation{Department of Astronomy, Tsinghua University, Beijing 100084, China}

\author[0000-0002-1301-3893]{Yi Mao}
\affiliation{Department of Astronomy, Tsinghua University, Beijing 100084, China}

\author{Cheng Cheng}
\affiliation{School of Chemistry and Physics, University of KwaZulu-Natal, Westville Campus, Durban, 4000, South Africa}

\author[0000-0002-5854-8269]{Benjamin D. Wandelt}
\affiliation{Sorbonne Universit\'e, CNRS, UMR 7095, Institut d'Astrophysique de Paris (IAP), 98 bis bd Arago, 75014 Paris, France}
\affiliation{Sorbonne Universit\'e, Institut Lagrange de Paris (ILP), 98 bis bd Arago, 75014 Paris, France}
\affiliation{Center for Computational Astrophysics, Flatiron Institute, 162 5th Avenue, New York, NY 10010, USA}



\begin{abstract}

Tomographic three-dimensional 21~cm images from the epoch of reionization contain a wealth of information about the reionization of the intergalactic medium by astrophysical sources. Conventional power spectrum analysis cannot exploit the full information in the 21~cm data because the 21~cm signal is highly non-Gaussian due to reionization patchiness. We perform a Bayesian inference of the reionization parameters where the likelihood is implicitly defined through forward simulations using density estimation likelihood-free inference (DELFI). We  adopt a trained 3D Convolutional Neural Network (CNN) to compress the 3D image data into informative summaries (DELFI-3D CNN). We show that this method recovers accurate posterior distributions for the reionization parameters. Our approach outperforms earlier analysis based on two-dimensional 21~cm images. 
In contrast, an MCMC analysis of the 3D lightcone-based 21~cm power spectrum alone and using a standard explicit likelihood approximation results in less accurate credible parameter regions than inferred by the DELFI-3D CNN, both in terms of the location and shape of the contours. Our proof-of-concept study implies that the DELFI-3D CNN can effectively exploit more information in the 3D 21~cm images than a 2D CNN or power spectrum analysis. This technique can be readily extended to include realistic effects and is therefore a promising approach for the scientific interpretation of future 21~cm observation data.

\end{abstract}

\keywords{Reionization (1383), H I line emission (690), Convolutional neural networks (1938)}


\section{Introduction}

The epoch of reionization (EoR), which marks the formation of first luminous objects, is one of the milestones in the evolution of our universe. During the EoR, the neutral hydrogen (H~{\small I}) gas in the intergalactic medium (IGM) is heated and ionized by ultraviolet and X-ray photons from the first luminous objects (see, e.g. \citealt{2018PhR...780....1D,2021MNRAS.503.3698H}). While the observations of high-redshift quasar spectra \citep[e.g.,][]{2006AJ....132..117F, 2015MNRAS.447..499M, 2015MNRAS.447.3402B} and the electron scattering optical depth to the cosmic microwave background (CMB) 
\citep{2020A&A...641A...6P} have placed robust observational constraints \citep{2015ApJ...811..140B,2015ApJ...802L..19R,2019ApJ...879...36F} on the EoR, the 21~cm line associated with the spin-flip transition of H~{\small I} atoms will be the most promising probe to cosmic reionization, because, in principle, the tomographic 21~cm survey contains the full three-dimensional (3D) information of H~{\small I} gas in the EoR, thereby directly revealing the astrophysical processes regarding how H~{\small I} gas in the IGM was heated and reionized by those ionizing sources. 

The next decades will be a golden age for 21~cm observations with a number of ongoing and upcoming experiments. Current interferometric arrays, including the Precision Array for Probing the Epoch of Reionization (PAPER, \citealp{Parsons2010}), the Murchison Wide field Array (MWA, \citealp{Tingay2013}), the LOw Frequency Array (LOFAR, \citealp{Haarlem2013}), and the Giant Metrewave Radio Telescope (GMRT, \citealp{2017A&A...598A..78I}), have first attempted to put upper limits on the 21~cm power spectrum from the EoR \citep{2013MNRAS.433..639P,2015ApJ...809...62P,2020MNRAS.493.1662M,2020MNRAS.493.4711T}. 
In the near future, the Hydrogen Epoch of Reionization Array (HERA, \citealp{DeBoer2017}) and the Square Kilometre Array (SKA, \citealp{Mellema2013}) promise to measure the 21~cm power spectrum from the EoR for the first time. More importantly, it is very likely that the SKA will have enough sensitivity to make 3D maps of the 21~cm signal. 

Unlike the CMB, the 21~cm signal is highly non-Gaussian, because patchy, bubble-like, structures of ionized hydrogen (H~{\small II}) regions are produced surrounding the ionizing sources. Thus, there is potentially a wealth of information in the 21~cm signal that is not contained in the 21~cm power spectrum, a two-point statistics of 21~cm brightness temperature fluctuations that is traditionally well studied in the literature. It is therefore essential to develop new methods that maximally exploit the full information in the 3D 21~cm images obtained by the SKA. Towards this goal, conventionally, new summary statistics that can only be measured with imaging have been proposed. These include the three-point correlation function \citep{2019MNRAS.487.3050H,2020MNRAS.498.4518J}, bispectrum \citep{2015MNRAS.451..266Y,2016MNRAS.458.3003S,2017MNRAS.468.1542S,2018MNRAS.476.4007M,2020MNRAS.499.5090M,2020MNRAS.497.2941S,2020MNRAS.492..653H,2021MNRAS.502.3800K}, one-point statistics \citep{2009MNRAS.393.1449H,2015MNRAS.451..467S,2021arXiv210204352G}, topological quantities such as the Minkowski functionals \citep{2006MNRAS.370.1329G,2019ApJ...885...23C,2021arXiv210103962K} and Betti numbers \citep{2020arXiv201212908G}, the cross-correlation between the 21~cm line and other probes such as the CO line \citep{2011ApJ...728L..46G,2011ApJ...741...70L}, the C~{\small II} line \citep{2012ApJ...745...49G,2018ApJ...867...26B}, the kinetic Sunyaev-Zel'dovich (kSZ) effect \citep{2018MNRAS.476.4025M,2020ApJ...899...40L}, and novel techniques such as the antisymmetric cross-correlation between the 21~cm line and CO line \citep{2021ApJ...909...51Z}. Since those summary statistics are fully determined by the parameters in the reionization models (hereafter ``reionization parameters''), in principle, Monte Carlo Markov Chain (MCMC) methods can be employed to constrain the reionization parameters from measurements of those statistics with futuristic 21~cm experiments (see, e.g. \citealt{2021arXiv210202310W}), just as the MCMC analysis with the 21~cm power spectrum \citep{2015MNRAS.449.4246G,2017MNRAS.472.2651G,Greig2018}. 

Ideally, an approach that could constrain the reionization parameters directly from the full 3D 21~cm images would probably have no loss of information. In principle, this could be done by writing down the non-Gaussian likelihood of the 21~cm images as a function of the reionization parameters, similar to the MCMC approach for analyzing the distribution of galaxies \citep{2013MNRAS.432..894J}. Nevertheless, such an explicit likelihood approach is a major effort and very computationally demanding.

Recently, machine learning has been extensively applied to 21~cm cosmology (\citealp{Shimabukuro2017, Kern2017Emulating, 2018anms.conf....7H,hassan2019constraining, Schmit2018,Jennings2019, jensen2016machine, doussot2019improved,Li2019,2020MNRAS.493.5913L,2021ApJ...907...44V,2021MNRAS.501.1463K,Gillet2019} hereafter referred to as \citetalias{Gillet2019}). In particular, \citetalias{Gillet2019} demonstrated that the information in the two-dimensional (2D) 21~cm image slices can be exploited to constrain the parameters for reionization and cosmic dawn with the 2D convolutional neural networks (CNN). However, they found that their constraint performance was just  comparable to the MCMC analysis with the 21~cm power spectrum. This can be explained by the fact that the 2D images contain only a subset of information in the 3D images. Nevertheless, this opens a new window to constrain reionization parameters, and even cosmological parameters \citep{hassan2019constraining},  using 21~cm images directly with machine learning techniques. Our paper is motivated to extend the work of \citetalias{Gillet2019} and demonstrate the applicability of the 3D CNN technique \citep{Baccouche2011,Cicek2016,Amidi2018} to the 21~cm 3D images for constraining reionization parameters. 

\begin{figure*}
	\includegraphics[width=\textwidth]{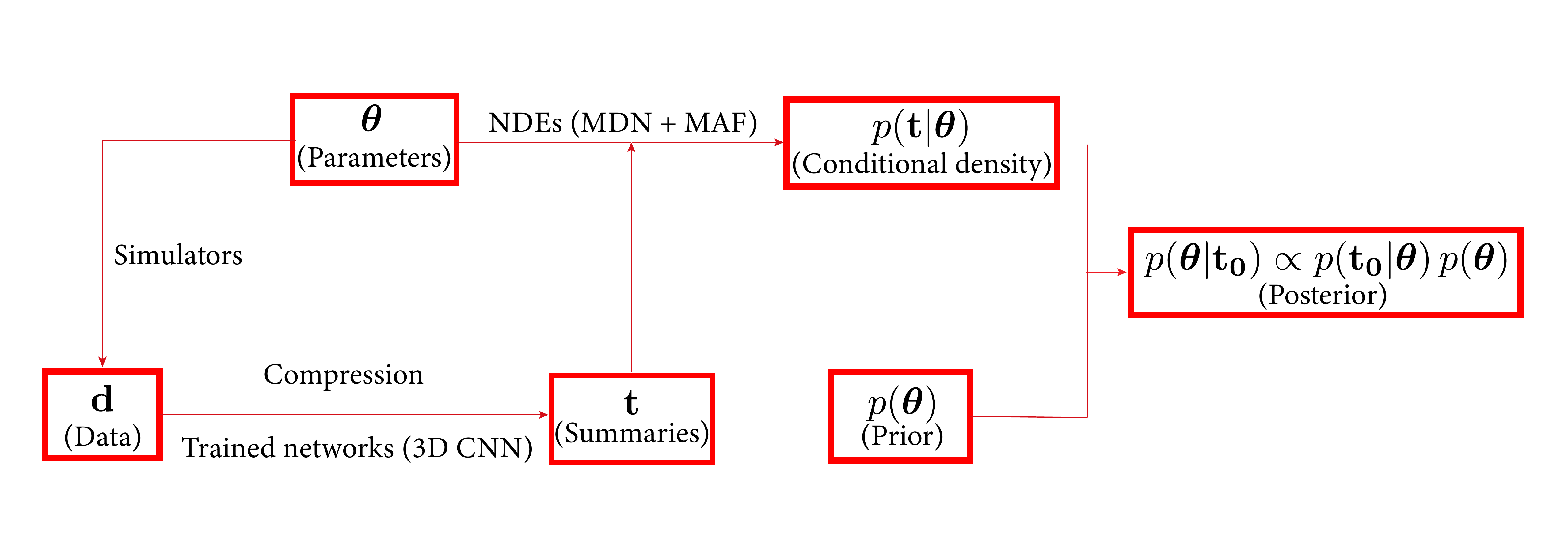}
    \caption{The workflow of DELFI-3D CNN. The data $\mathbf{d}$ is prepared using the simulator ({\tt 21cmFAST} herein) with the parameters $\boldsymbol{\theta}$. The data is compressed to the low-dimensional summaries $\mathbf{t}$ using some compressors (trained 3D CNN herein). The parameter-summary pairs ($\boldsymbol{\theta}, \mathbf{t}$) are fed to an ensemble of NDEs that are trained to learn the data likelihood $p(\mathbf{t} | \boldsymbol{\theta})$. The posterior distribution is inferred from the learned data likelihood and parameter prior using  Bayes' Theorem.}
    \label{fig:pro}
\end{figure*}

It is worthwhile noting that most 21~cm cosmology literature using machine learning did not include the posterior inference for recovered parameters (not to be confused with the estimation of recovery errors with respect to the true parameter values). In this regard, \citet{2020arXiv200507694H,H2020Parameters} applied the Bayesian neural networks (BNNs, \citealp{2015Dropout,levasseur2017uncertainties,2019Parameters}) to the 21~cm 2D images, giving an estimate of reionization parameters and a measure of the uncertainties. Ideally, BNNs can infer a posterior distribution over the network weights and output principled expectation values and uncertainties assuming a simple multivariate Gaussian distribution. For this purpose, various approximations were made in the literature to realize the BNNs. However, the inconsistency between different approaches is still a problem \citep{2020arXiv200507694H,H2020Parameters,2019Parameters}. 

Instead, in this paper, we employ the density-estimation likelihood-free inference (DELFI, \citealp{bonassi2011bayesian,fan2013approximate,papamakarios2016fast,lueckmann2017flexible,Kern2017Emulating,alsing2018massive,lueckmann2018likelihood,2020MNRAS.tmp.2709R,alsing2019fast} hereafter referred to as \citetalias{alsing2019fast}), a more flexible framework than the CNN to give the posterior inference, including both parameter estimation and Bayesian uncertainty estimation. Unlike  Bayesian inference based on an explicit likelihood \citep{2015MNRAS.449.4246G,2017MNRAS.472.2651G,Greig2018,Park2018}, the so-called likelihood-free inference (LFI) defines the likelihood implicitly through forward simulations. This allows building a sophisticated data model without relying on approximate likelihood assumptions. Instead, DELFI contains various neural density estimators (NDEs) to learn the likelihood as the conditional density distribution of the target data given the parameters from a number of simulated parameter-data pairs. It has been demonstrated to need fewer model simulations to get the high-fidelity posterior distribution than the traditional Approximate Bayesian Computation (ABC, \citealp{schafer2012likelihood,cameron2012approximate,weyant2013likelihood,robin2014constraining,ishida2015cosmoabc,lin2015new,carassou2017inferring,kacprzak2018accelerating,akeret2015approximate,davies2018new,ishida2015cosmoabc,hahn2017approximate}) when comparing the convergence speed. In comparison with the BNNs, the optimization of an ensemble of networks in the DELFI is typically simpler and cheaper than the posterior inference over the large number of weights of a single large network in the BNNs. In addition, it is easy for DELFI to average different network architectures by taking advantage of neural ensembles
\citepalias{alsing2019fast}. 

In this paper, we first train the 3D CNN which outputs the predicted values of reionization parameters from the 21~cm lightcone 3D images. While these estimates have physical meaning, note that from the DELFI point of view, these parameter values predicted from the 3D CNN are data {\it summaries} of the input 3D images, and the 3D CNN is just a data {\it compressor}. DELFI itself is a framework of posterior inference to provide both parameter estimation (technically, independent of the 3D CNN prediction) and statistical uncertainty estimation. We then apply a public implementation of DELFI with NDEs, {\tt pydelfi}\footnote{https://github.com/justinalsing/pydelfi} \citepalias{alsing2019fast}, to perform the posterior inference. 
As a demonstration of concept, we simplify our simulations of mock samples in two aspects. First, while both reionization parameters and cosmic dawn astrophysical parameters were constrained in \citetalias{Gillet2019}, our simulations are restricted to the regime when the 21~cm spin temperature is much larger than the CMB temperature, so the contributions from the cosmic dawn astrophysical parameters are negligible. This assumption holds well after reionization begins. Secondly, as in \citetalias{Gillet2019}, our simulations do not include the effects of thermal noise and residual foregrounds. Since the DELFI framework is flexible enough, these effects can be readily taken into account in the future work. Our paper improves upon \citetalias{Gillet2019} in exploiting the 3D 21~cm images with 3D CNN, and performing a self-consistent posterior inference with {\tt pydelfi}. 

The rest of this paper is organized as follows. In Section~\ref{sec:DELFI}, we introduce our integrated framework of DELFI with 3D CNN as a data compressor (hereafter ``DELFI-3D CNN''). We give the results of DELFI-3D CNN in Section~\ref{sec:results}, and make concluding remarks in Section~\ref{sec:conc}. Some technical discussions are left to Appendix~\ref{sec:opt} (on the optimization of 3D CNN), Appendix~\ref{app:hypo} (on the effect of missing ${\bf k}_{\perp} = 0$ mode in the interferometric measurement), Appendix~\ref{app:ndemath} (on the formalism of NDEs), Appendix~\ref{ICs} (on the effect of initial conditions), Appendix~\ref{app:stack} (on the stabilization of different NDEs), and Appendix~\ref{app:convergence} (on the convergence of the {\tt 21CMMC} code). 

\begin{figure*}
	\includegraphics[width=\textwidth]{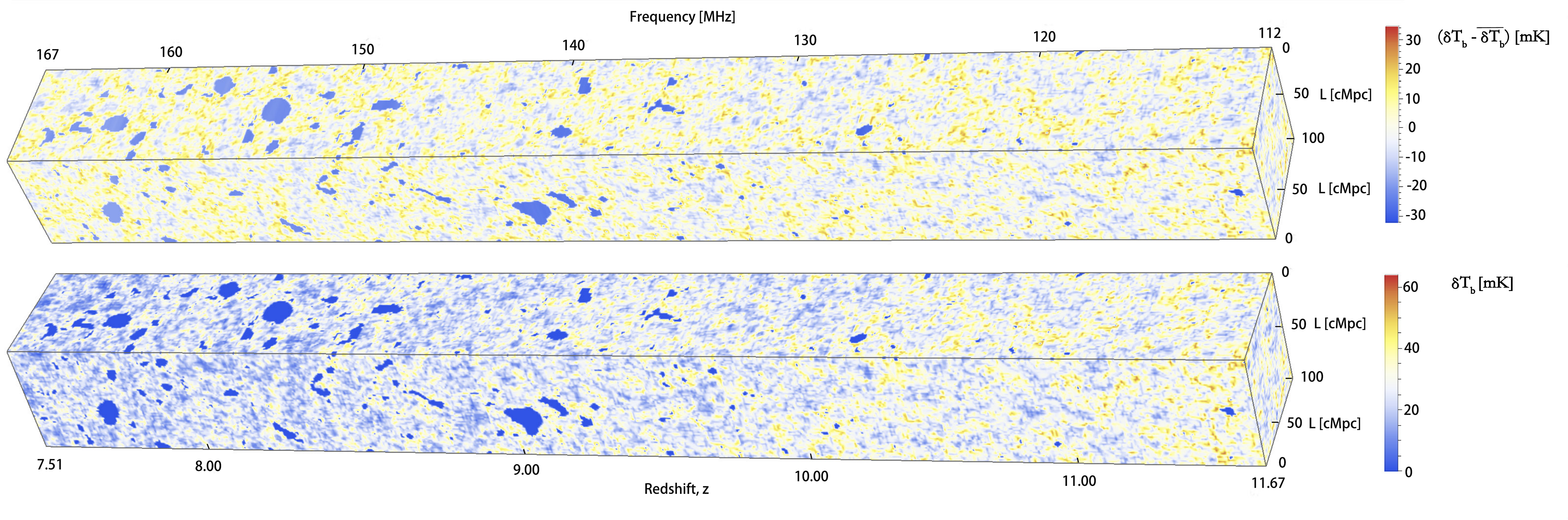}
    \caption{An illustration of the lightcone datacube of 3D 21~cm images $\delta T_{\rm b}$ (bottom) in the redshift range $z=7.51 - 11.67$. To mimic the observations from radio interferometers, we subtract from the lightcone field the mean of the 2D slice for each 2D slice perpendicular to the line-of-sight, and form the observed 3D 21~cm images (top), $\delta T_{\rm b} - \overline{\delta T_{\rm b}}$.}
    \label{fig:sample}
\end{figure*}

\section{DELFI-3D CNN Methodology}
\label{sec:DELFI}
DELFI transforms the problem of parameter inference into density estimation with the simulated parameter-data pairs. To implement DELFI, we adopt a flexible and effective approach, with the workflow illustrated in Fig.~\ref{fig:pro} --- we first learn the conditional density $p(\mathbf{t} | \boldsymbol{\theta})$ by maximizing the total data likelihood, where $\boldsymbol{\theta}$ and $\mathbf{t}$ are the parameter vector and data summary vector, respectively, and then infer the posterior using the Bayes' Theorem, $p(\boldsymbol{\theta} | \mathbf{t_0}) \propto p(\mathbf{t_0} | \boldsymbol{\theta}) \, p(\boldsymbol{\theta})$ at any data summary $\mathbf{t_0}$ from observed data, where $p(\boldsymbol{\theta})$ is the prior. In view of large-scale data from simulations, we employ the 3D CNN to compress the 21~cm 3D images (the data $\mathbf{d}$) into the data summaries $\mathbf{t}$, and train the DELFI with a large number of parameter-summary pairs ($\boldsymbol{\theta}, \mathbf{t}$). In this section, we introduce our methodology in detail.

\subsection{Data Preparation}
\label{sec:data}
In this paper, we use the publicly available code {\tt 21cmFAST}\footnote{https://github.com/andreimesinger/21cmFAST} \citep{Mesinger2007,Mesinger2011}, which can be used to perform semi-numerical simulations of reionization, as the simulator to generate the datasets. This code uses the excursion-set approach \citep{2004ApJ...613....1F} to identify ionized regions. It quickly generates the fields of density, velocity, ionization field, spin temperature and 21~cm brightness temperature on a grid.

Our simulations were performed on a cubic box of 100 comoving ${\rm Mpc}$ on each side, with $66^3$ grid cells. The 21~cm brightness temperature at position ${\bf x}$ relative to the CMB temperature can be written \citep{Furlanetto2006} as 
\begin{equation}
T_{21}(\textbf{x},z)=\tilde{T}_{21}(z)\,x_{\rm HI}(\textbf{x})\,\left[1+\delta(\textbf{x})\right]\,(1-\frac{T_{\rm CMB}}{T_S})\,,\label{eqn:21cm}
\end{equation}
where $\tilde{T}_{21}(z) = 27\sqrt{[(1+z)/10](0.15/\Omega_{\rm m} h^2)}(\Omega_{\rm b} h^2/0.023)$ in units of mK. 
Here, $x_{\rm HI}({\bf x})$ is the neutral fraction, and $\delta({\bf x})$ is the matter overdensity, at position ${\bf x}$. We assume the baryon perturbation traces the cold
dark matter on large scales, so $\delta_{\rho_{\rm H}} = \delta$. 
In this paper, we focus on the limit where spin temperature $T_S \gg T_{\rm CMB}$, valid soon after reionization begins. As such, we can neglect the dependence on spin temperature. Also, for simplicity, we ignore the effect of peculiar velocity, because it only weakly affects the light-cone effect. 

We generate a realization of the 21~cm brightness temperature fields using the density and ionized fraction fields from semi-numerical simulations with the code {\tt 21cmFAST}, given an initial condition in the density fields. We then interpolate the snapshots at different time to construct the lightcone data cube along the line-of-sight (LOS) within a comoving distance of a simulation box. To reduce the interpolation error caused by insufficient sampling of snapshots, we output the simulation results at 9 different redshifts within the corresponding cosmic time. To avoid the impact of periodic boundary condition, which otherwise results in the repeated structures along the LOS due to the same initial density fields, we concatenate ten such lightcone boxes, each simulated with different initial conditions in density fields but with the same reionization parameters, together to form a full lightcone datacube of the size $100\times 100\times 1000$ comoving ${\rm Mpc}^3$. To mimic the observations from radio interferometers, we subtract from the lightcone field the mean of the 2D slice for each 2D slice perpendicular to the LOS, because radio interferometers cannot measure the mode with ${\bf k}_\perp =0 $.\footnote{Note that \citetalias{Gillet2019} did not implement this step of subtracting the mean of the 2D slice from the 21~cm signal. The effect of missing the ${\bf k}_\perp =0 $ mode in the interferometer measurement is discussed in Appendix~\ref{app:hypo}.} This forms the mock lightcone datacube of $\delta T_{\rm b} - \overline{\delta T_{\rm b}}$ in the redshift range $ 7.51 \le z \le 11.67$ (see Fig.~\ref{fig:sample}). 

Our reionization model is parametrized with two parameters as follows:

(1) $\zeta$, the {\it ionizing efficiency}. $\zeta=f_{\rm esc}f_{*}N_{\gamma}/(1+\overline{n}_{\rm rec})$ \citep{Furlanetto2004,Furlanetto2006}, which is a combination of several parameters related to ionizing photons.  Here, $f_{\rm esc}$ is the fraction of ionizing photons escaping from galaxies into the IGM, $f_{*}$ is the fraction of baryons locked in stars, $N_{\gamma}$ is the number of ionizing photons produced per baryon in stars, and $\overline{n}_{\rm rec}$ is the mean recombination rate per baryon. In our dataset, we vary $\zeta$ in the range of $10\le \zeta \le 250 $.

(2) $T_{{\rm vir}}$, the {\it minimum virial temperature of haloes that host ionizing sources}. Typically, $T_{\rm vir}$ is about $10^{4} {\rm K}$, corresponding to the temperature above which atomic cooling becomes effective. In our dataset, we explore the range of $10^{4} \le T_{{\rm vir}} \le 10^{6}\,{\rm K}$. 

These ranges are consistent with \citet{Greig2018}. Other parameters that have less impact on reionization are fixed throughout this paper, e.g. $R_{\rm mfp}=15\,{\rm Mpc}$ (the mean free path of ionizing photons). Also, cosmological parameters are fixed as $\mathrm{( \Omega _ { \Lambda } , \Omega _ { m } , \Omega _ { b } , n_s , \sigma _ { 8 }} , h )=( 0.692,0.308,0.0484,0.968,0.815,0.678 )$ \citep{ade2016planck}. 

For each given set of reionization parameters, we generate a mock lightcone datacube of 3D 21~cm images, or a sample. We use the Latin Hypercube Sampling \citep{10.2307/1268522} to scan the EoR parameter space. For the 3D CNN, we generate 10,000 samples, in which 8,000 samples are used for training the 3D CNN, 1,000 for validation, and 1,000 for testing the 3D CNN. 
After the 3D CNN is trained, its network weights are fixed. Then we generate 10,000 new samples which are compressed into parameter-data summaries pairs by the trained 3D CNN, and then provided to the DELFI for training the density estimators (with 8,100 samples), validation (with 900 samples), and testing the DELFI-3D CNN (with 1,000 samples). 
The initial conditions for all realizations were independently generated by sampling spatially correlated Gaussian random fields with the power spectrum given by linear theory.\footnote{The effect of initial conditions on the parameter estimation is discussed in Appendix~\ref{ICs}.}

\begin{figure*}
	\includegraphics[width=\textwidth]{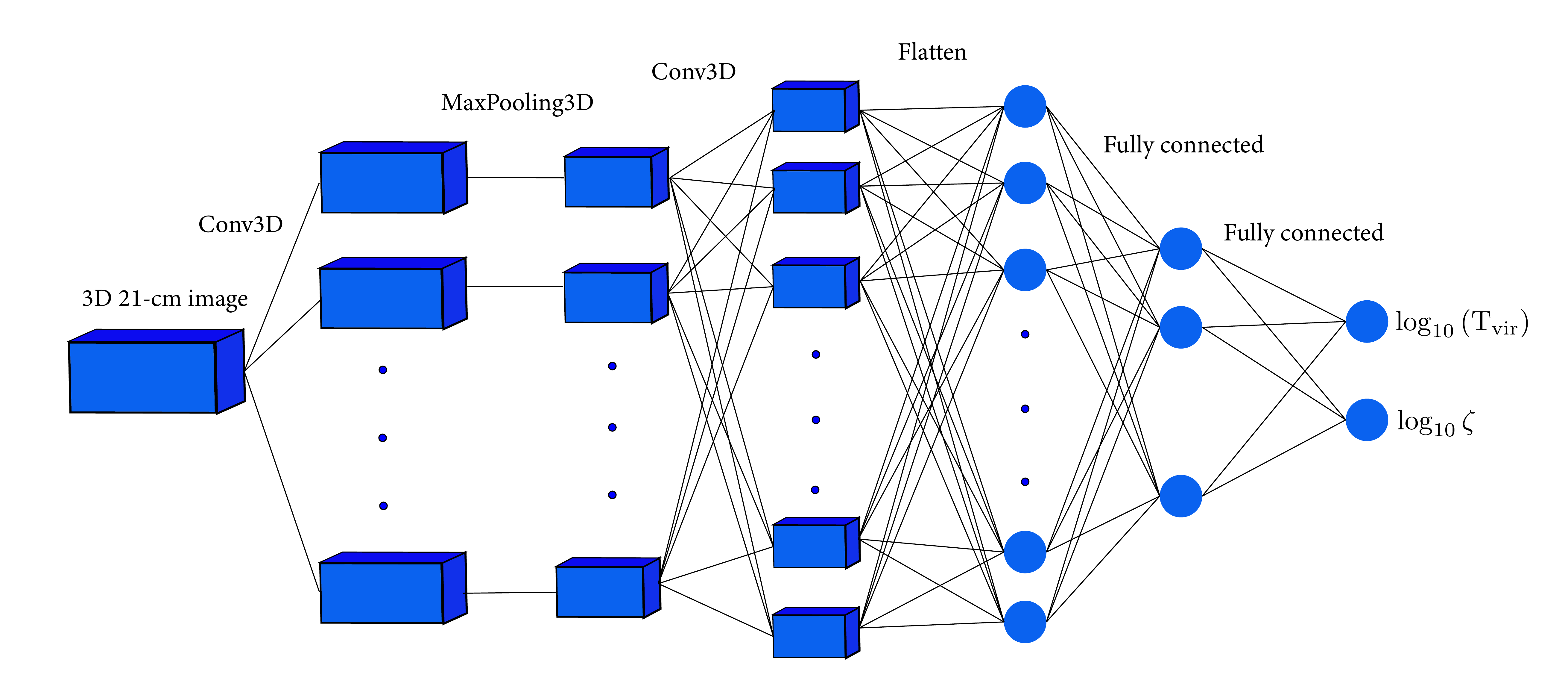}
    \caption{An illustration of main structure in the 3D CNN. The input set is the 3D 21~cm images, and the output is a set of two reionization parameters, $\log_{10}(T_ { \rm vir })$ and $\log_{10}(\zeta)$. Here we show the main 3D Convolution (Conv3D), Maxpooling3D, Flatten, and fully connected layers.}
    \label{fig:stru}
\end{figure*}

\subsection{3D CNN}
\label{sec:structure}

We train a 3D CNN to get an estimator of the reionization parameters from the 3D 21~cm lightcone datacube.  From the DELFI perspective, the trained 3D CNN is a compressor that compresses the 3D 21~cm images (the data $\mathbf{d}$) into the estimates of reionization parameters (the summaries $\mathbf{t}$). In this subsection, we describe in detail the structure of a typical 3D CNN, the step-by-step process of feature extraction in 3D CNN, the training strategy, and some practical aspects of hyper-parameter tuning.   

\subsubsection{Structure of 3D CNN}
Basically, a 3D CNN can be set up by replacing the 2D operations with 3D counterparts, including 3D Convolution (Conv3D), 3D MaxPooling, 3D ZeroPadding, and 3D Dropout. We illustrate the main structure in 3D CNN in Fig.~\ref{fig:stru}. Interested readers are referred to \citetalias{Gillet2019} for the 2D CNN that was applied to astrophysical parameter estimation. 

A network can be built by stacking many layers together. When the input image goes through the convolutional operations, the image features are extracted by some 3D kernels each with certain size. Each kernel sweeps through the entire image with specific stride size in a specific order. After that, one corresponding feature map is generated. A ZeroPadding operation before the Convolution operation can be used to offset the shape size decrease in the output image from the former layer. A BatchNormalization operation \citep{Ioffe2015a,10.5555/3327144.3327174} can follow a Convolution one to enable faster and more stable training. The batch normalization first normalizes the input to have the expectation value of 0 and the variance of 1, and then introduces a linear transform to the normalized values to ensure identity transform \citep{Ioffe2015a}. It is usually applied to mini-batches, which are drawn from the whole training data. Non-linear functions like Rectified Linear Units (ReLU) \citep{dahl2013improving}, $f(x)=\max(0,x)$, are applied to enable the network to create complex mappings between the inputs and outputs. After the Convolution layer, a Pooling operation, which serves as a down-sampling strategy, can reduce the number of parameters and learn position invariant features. In the fully connected layer, neurons have full connections to all activations in the former layer. The Dropout operation \citep{JMLR:v15:srivastava14a,NIPS2013_71f6278d} can be applied to tackle potential over-fitting of the network by randomly dropping units from the neural network during training, preventing units from co-adapting too much. Finally, the output of the network can be either class labels for classification problems or continuous variables for regression problems. The parameter estimation is a typical example of a regression problem.

\begin{table}
	\centering
	\caption{The stacked layers with their output shapes in the 3D CNN.}
	\begin{tabular}{lcr} 
		\hline\hline
		Step & Layer type & Output shape\\
		\hline
		1 & Input & $66\times66\times660$\\
		2 & $5\times5\times5$ Conv3D & $31\times31\times328$\\
		3 & BatchNormalization+ReLU & $31\times 31\times 328$\\
		4 & $2\times2\times2$ MaxPooling3D & $15\times 15\times 164$\\
		5 & ZeroPadding3D & $17\times 17\times 166$\\
		6 & $5\times5\times5$ Conv3D & $13\times 13\times 162$\\
		7 & BatchNormalization+ReLU & $13\times 13\times 162$\\
		8 & $2\times2\times2$ MaxPooling3D & $6\times 6\times 81$\\
		9 & SpatialDropout3D(40\%) & $6\times 6\times 81$\\
		10 & ZeroPadding3D & $8\times 8\times 83$\\
		11 & $5\times5\times5$ Conv3D & $4\times 4\times 79$\\
		12 & BatchNormalization+ReLU & $4\times 4\times 79$\\
		13 & $2\times2\times2$ MaxPooling3D & $2\times 2\times 39$\\
		14 & SpatialDropout3D(40\%) & $2\times 2\times 39$\\
		15 & Flatten & $19968$\\
		16 & Fully connected & 64\\
		17 & BatchNormalization+ReLU & 64\\
		18 & Dropout(40\%) & 64\\
		19 & Fully connected & 16\\
		20 & BatchNormalization+ReLU & 16\\
		21 & Fully connected & 4\\
		22 & BatchNormalization+ReLU & 4\\
		23 & Fully connected Left& 1\\
		24 & Fully connected Right& 1\\
		\hline
	\end{tabular}
	\label{tab:struc}
\end{table}

\subsubsection{Network Training}

For regression problems, loss functions like the mean absolute error (MAE) and mean squared error (MSE) aim to penalize the deviation between the true and predicted values. Our objective is to attain a loss minimum where the loss function reaches the smallest possible value, which may be facilitated by using a proper optimizer like {\tt RMSprop} \citep{tieleman2012lecture}. The key ingredient for an optimizer is the recipe for an appropriate learning rate: too large a learning rate often leads to local minima, but  convergence is hard to achieve if the learning rate is too small. In this paper, we choose a variable learning rate that decreases by a factor of 10 once the learning begins to stagnate. 

Generally, a validation dataset that is independent of the training and test datasets is used during the training process, for two purposes --- to choose the optimal model on the validation dataset, and avoid over-fitting while training. Over-fitting can be recognized if errors on the validation dataset increase. A strategy for faster convergence to an optimal model is to set early stopping, which ends the training when errors on the validation dataset stop decreasing. Once that happens, we can choose a new set of hyper-parameters for training. In view of the GPU memory limitation, we can split the training dataset into ``mini-batches''. After processing one mini-batch of training data, the network updates the network weights through back-propagation, which are the actual parameters trained throughout the network. Note that inefficient computation occurs (due to noisy stochastic gradients) when the size of mini-batch is too small, so the mini-batch size serves as a trade-off between the memory limitation and computational efficiency. We choose a mini-batch size of 32 in this paper. 

We adopt a single 3D CNN model that is however flexible enough. The network setup includes one Convolutional layer with filter size of $5 \times 5 \times 5$, one MaxPooling layer, two blocks (each containing one ZeroPadding layer, one Convolutional layer, one MaxPooling layer and one 3D Dropout layer), the Flatten layer, one fully-connected layer with 64 neurons, one 1D Dropout layer, and two successive fully-connected layers with 16 and 4 neurons, respectively. The dropout rate for all kinds of Dropout layers is 0.4. The total number of parameters to be trained is about 2.56 million. The layer types and their output shapes are listed in Table~\ref{tab:struc}. 
We use the Keras functional API\footnote{https://keras.io/getting-started/functional-api-guide/} to set up the CNN model.

Due to the large dynamical range of reionization parameters, we choose to output these parameters in the logarithmic scale, in the default range ($4 \le \log_{10}(T_{{\rm vir}}/{\rm K}) \le 6$, $ 1\le \log_{10}(\zeta) \le 2.398 $). We also optimize the network with the output weights chosen to be 0.8 and 1.0 for $\log(T_{\rm vir}) $ and $\mathrm{log(\zeta)}$, respectively, in order to balance the recovery performance between these two parameters, because the 21~cm signals depend on them with slightly different sensitivity (see the 2D case in \citetalias{Gillet2019}). The recovery performance is roughly stabilized among different network setups, which we discuss in detail in Appendix~\ref{sec:opt}. The aforementioned setups, which we adopt throughout this paper unless noted otherwise, result from the optimization of the recovery performance. In this paper, we use 64~GB memory for loading and training data, and one NVIDIA GeForce GTX 1080~Ti GPU card with 11~GB RAM memory for computational speedup. It takes 6 minutes for each epoch of training, and a typical model training is finished within 70 epochs. After training, it takes 18~milliseconds for the trained 3D CNN to process a test sample.  

\subsection{Neural Density Estimators (NDEs)}
\label{sec:NDE}
NDEs can model the probability distribution (or density) conditioned on the parameters $p(\mathbf{t}|\mathbf{\theta})$. They are powerful tools for DELFI. Two implementations of NDEs have been proven successful --- mixture density networks (MDNs, \citealp{bishop1994mixture}), and masked autoregressive flows (MAFs, \citealp{papamakarios2017masked}). We give a brief summary in this subsection and leave the technical details to Appendix~\ref{app:ndemath}, based on the work of \citet{bishop1994mixture}, \citet{germain2015made}, \citet{papamakarios2017masked}, \citet{JMLR:v22:19-1028}, and \citetalias{alsing2019fast}.

\subsubsection{Mixture density networks (MDNs)}

MDNs combine typical neural networks with a mixture density model such as Gaussian density. It can represent more complexity of the conditional density when more mixture components are contained. With enough Gaussian components, in principle, MDNs can model the underlying non-Gaussian properties in the conditional data density.

\subsubsection{Masked autoregressive flows (MAFs)}

The word ``flows'' here refers to the normalizing flows \citep{JMLR:v22:19-1028}, which transform a simple base density with a series of transformations to a richer distribution, in analogy to a fluid flowing through a set of tubes. The base density is often taken to be a multivariate normal, so the inverse transformation from any distribution back to the base density is called ``normalizing''.

The transformation is required to be invertible and both the forward transformation and inverse transformation should be differentiable. These properties also ensure that the transformations are composible, which increases the actual expressive power of the flow-based models.

With a flow-based model, one can either draw samples from the model with the base density and the forward transformation, or evaluate the model's density with the inverse transform and its Jacobin determinants and the base density. These two options can be chosen for specific needs. In this paper we need to evaluate the model's density so we choose the latter one. 

The autoregressive model can be interpreted as a normalizing flow. For the autoregressive model, the density of vector $\mathbf{t}$ can be expressed as a product of one-dimensional density by the chain rules, where the autoregressive property means that the density of the component $t_i$ only depends on the previous $i-1$ components. The masked autoencoders for density estimation (MADE) provides an efficient way to implement the autoregressive property with a modified fully-connected auto-encoder. Consider a Gaussian distribution as the individual one-dimensional density, by learning the mean and variance of each density and using the exponential activation function to ensure the positivity, the MADE can be interpreted as the inverse transformation from the vector $\mathbf{t}$ back to a random vector following a standard normal distribution. The autoregressive property makes the Jacobin determinants tractable, which is feasible for density estimation.

In practice, we stack multiple MADEs to form the normalizing flows. The ordering of input data components to construct the autoregressive density can be varied for each MADE in order to increase the flexibility. In this paper, the MAFs are conditioned on parameter $\mathbf{\theta}$. This can be easily implemented by constructing the chain rule of densities conditioned on $\mathbf{\theta}$.

Training the NDEs (including both MAFs and MDNs) is basically fitting the density estimators $p(\mathbf{t} | \boldsymbol{\theta};\mathbf{w})$ with dependence on the network weights $\mathbf{w}$ to a target density $p^{\star}(\mathbf{t} | \boldsymbol{\theta})$ by minimizing some divergence, e.g.\ the Kullback-Leibler divergence, between them.

\subsubsection{NDE Setup}

Both MDNs and MAFs are included in the {\tt pydelfi} package. Training an ensemble of NDEs can help to get robust results on small training sets, as well as avoiding the risk of over-fitting \citepalias{alsing2019fast}. We construct herein the density estimator from an ensemble of 4 MDNs (each with 1, 4, 6, and 8 Gaussian components, respectively), and a MAF (with 8 MADEs). We checked and confirmed that other ensembles of networks give similar results (see Appendix~\ref{app:stack}). 

We train the NDEs with the batch size of 100, epochs of 500, and early stop patience of 20 (which means that the training is finished when there is no loss decreasing over 20 epochs).  

\subsection{Other Setups of DELFI}

For the input of DELFI, the trained 3D CNN with fixed weights compresses the 21~cm 3D images into the two-dimensional summaries (corresponding to two-dimensional reionization parameter space). 

The choice of prior in the parameters is flexible. If active learning is taken, e.g. in \citetalias{alsing2019fast}, the prior is given by the learning process. In our paper, we choose a flat prior $p(\boldsymbol{\theta})$ over the default region in parameter space, and produce the training set using parameters $\theta$ sampled from it before running DELFI.

To sample the posteriors, we run a Markov Chain Monte Carlo on the Bayesian posterior output by DELFI. We run 100 walkers for 3000 steps and drop the first 500 steps as  ``burn in,'' because the integrated autocorrelation time is estimated to be around 30. We have tested the results of posterior inference with a larger number (up to 100,000) of total steps, and find the convergence for $\ge 3000$ steps.

\begin{figure*}
	\includegraphics[width=\textwidth]{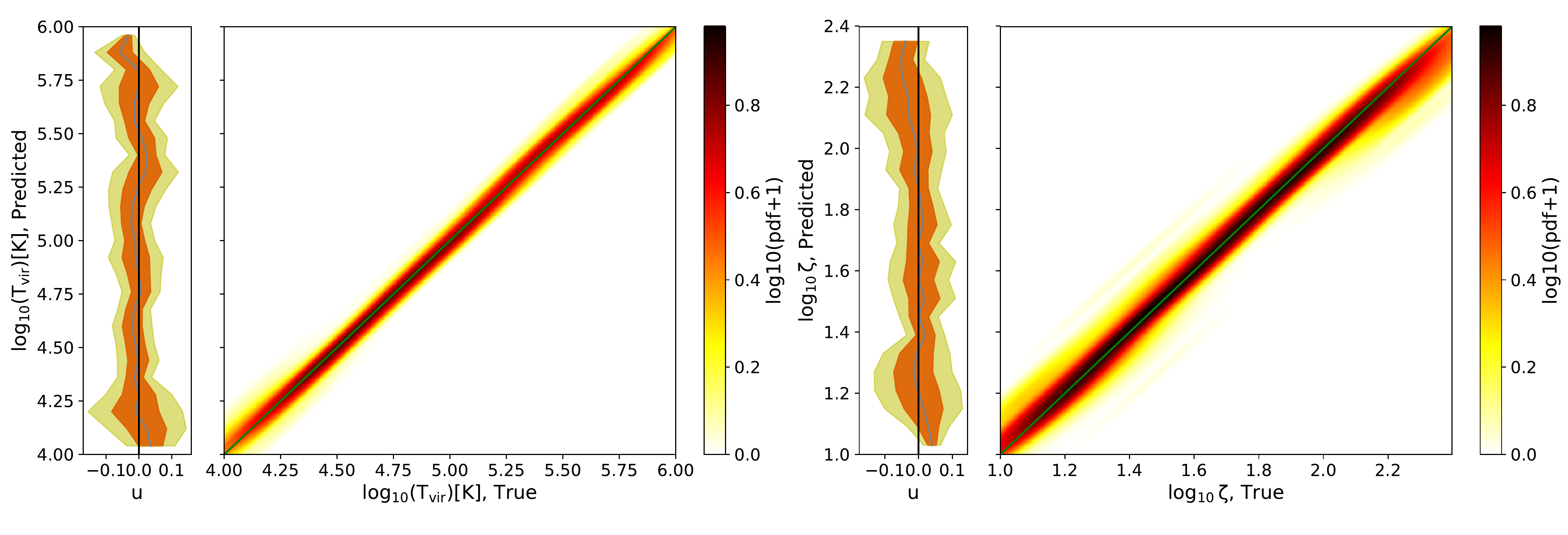}
    \caption{Reionization parameter recovery performance with 3D CNN using 1000 test samples. Shown are the predicted value $y_{\rm pred}$ vs the true value $y_{\rm true}$ of each parameter $y = \log_ { 10 }(T_{\rm vir})$ (left) and $\log_ { 10 }(\zeta)$ (right), respectively, with the color representing $\log_{10}(1+{\rm pdf})$ where pdf is the probability density function of sampled data points}. The green diagonal line indicates the perfect (zero-error) recovery. The side histograms show the distribution of residuals $u = y_{\rm pred} - y_{\rm true}$ at a given predicted value, with the mean (grey line) and 68\% (orange region) and 95\% (yellow region) probability intervals.  
    \label{fig:comp}
\end{figure*}

\begin{figure*}
	\includegraphics[width=\textwidth]{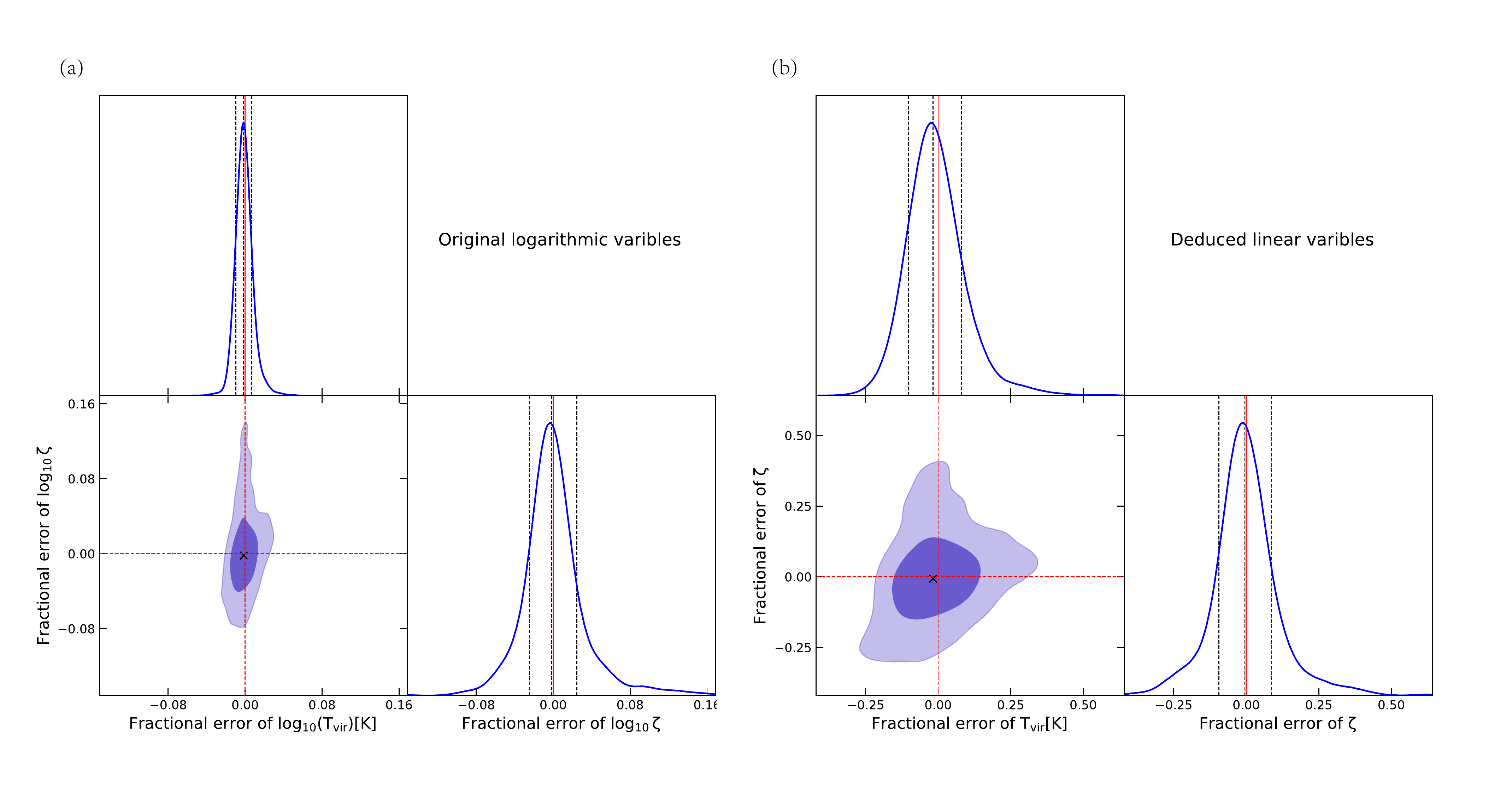}
    \caption{Reionization parameter recovery performance with 3D CNN using 1000 test samples. Shown are the 2D joint probability density distribution of the accuracy of the estimated parameters, i.e. the fractional error $\epsilon = (y_{\rm pred} - y_{\rm true})/y_{\rm true}$, where $y$ represents either the parameters directly predicted from the 3D CNN, $ \log(T_{\rm vir})$ and $\log(\zeta)$ (left), or the deduced parameters, $ T_{\rm vir}$ and $\zeta$ (right). 
    In the 1D marginalized distribution of each panel, the quantiles of 0.16, 0.5 and 0.84 are marked with black dash lines, while the red solid line indicates the perfect (zero-error) recovery. 
    In the 2D joint probability density distribution of each panel, we show the 68\% (dark purple region) and 95\% (light purple region) probability regions, the median (black cross), and the perfect (zero-error) recovery (red dash lines). }
    \label{fig:fractional error}
\end{figure*}

\begin{figure*}
	\includegraphics[width=\textwidth]{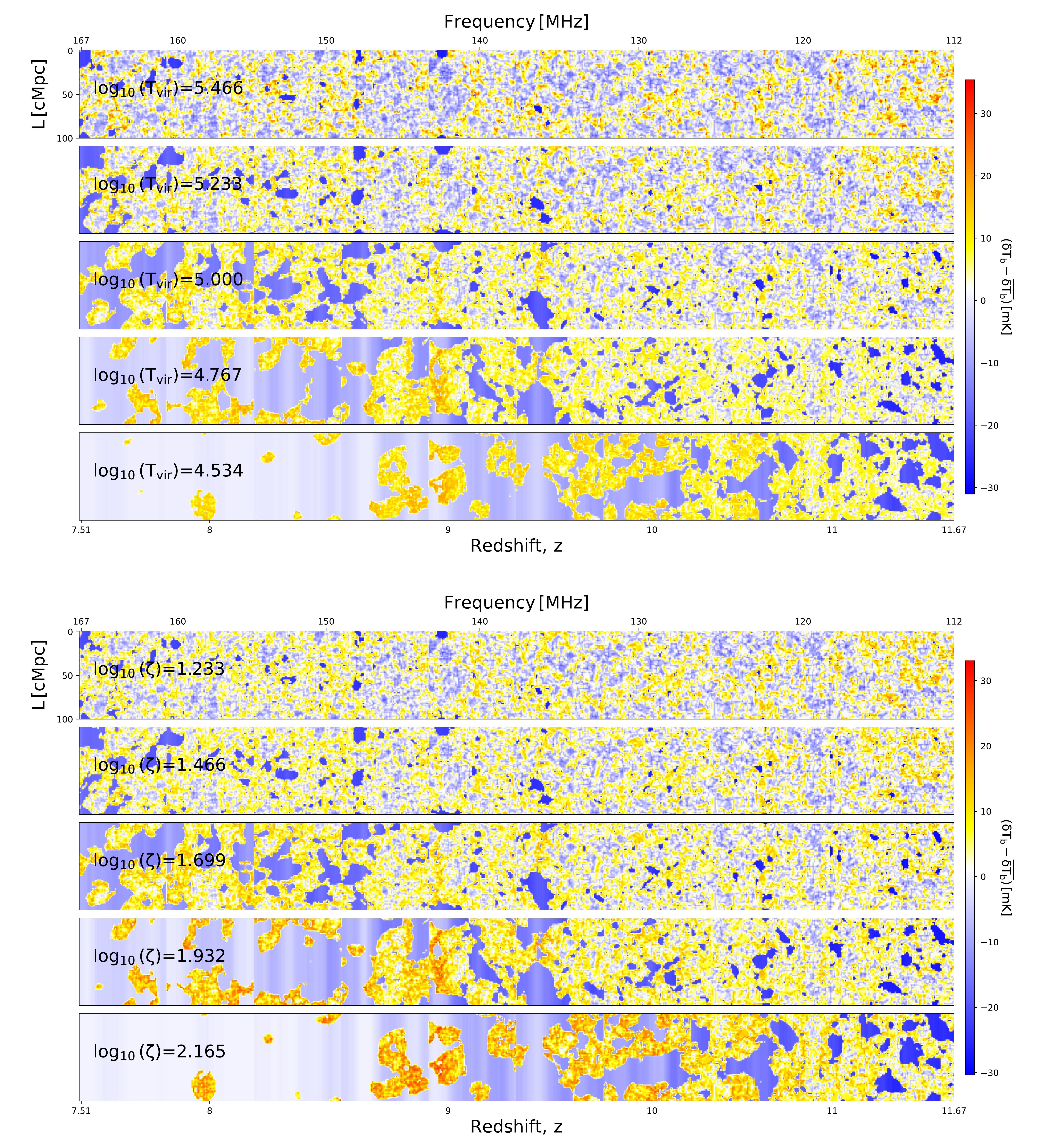}
    \caption{An illustration of the impact of varying parameters on the signal. (Top) we show the lightcone 21~cm images with fixed $\log _ { 10 } \left( \zeta\right)=1.699$ and varying $\log _ { 10 } \left( T_ { \rm vir }\right) = 5.466$, 5.233, 5.000, 4.767, 4.534, respectively. (Bottom) we show the lightcone 21~cm images with fixed $\log _ { 10 } \left( T_ { \rm vir }\right)=5.000$ and varying $\log _ { 10 } \left( \zeta\right) = 1.233$, 1.466, 1.699, 1.932, 2.165, respectively. The middle lightcones in both top and bottom panels are the same and used as the benchmark for comparison. }
    \label{fig:vary_params}
\end{figure*}

\begin{figure*}
	\includegraphics[width=\textwidth]{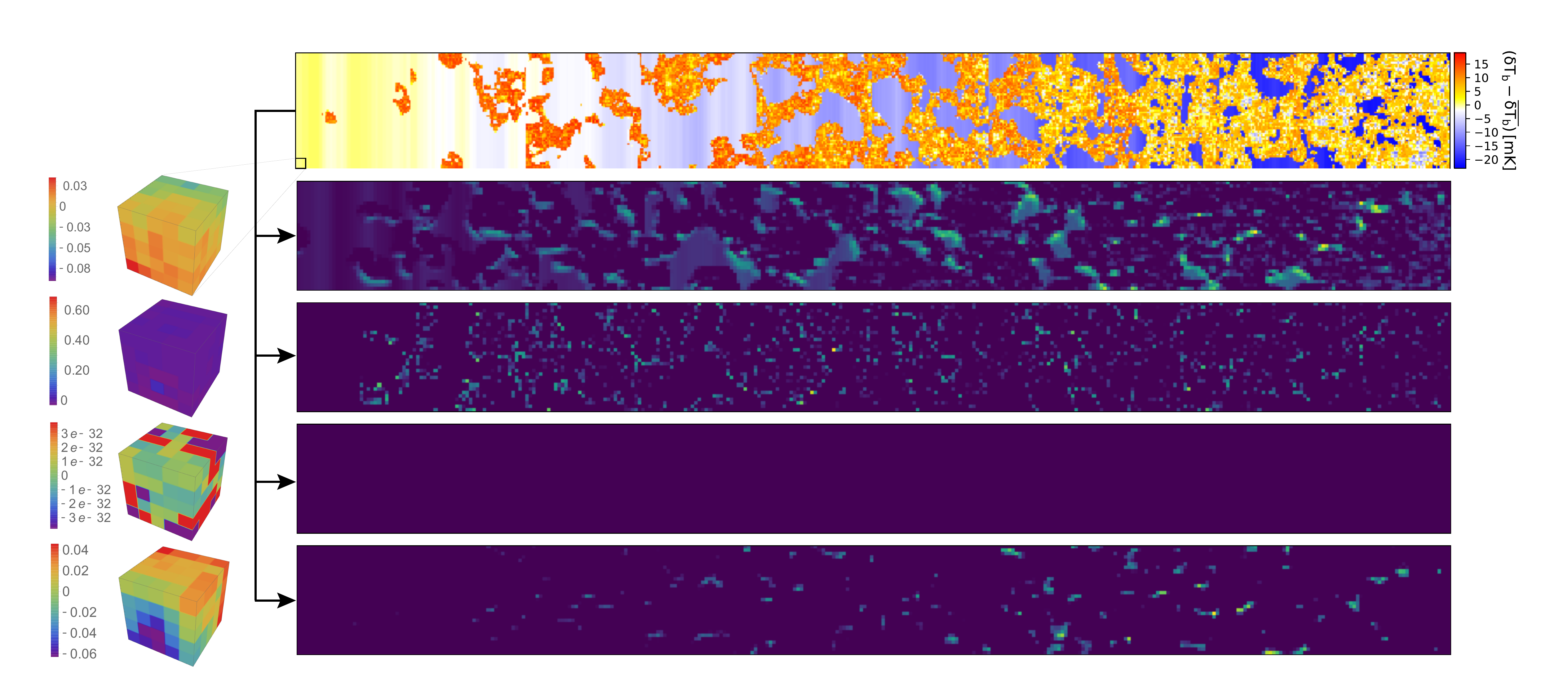}
    \caption{An illustration of kernels in 3D CNN. (Top) input lightcone 21~cm images in $66\times 66 \times 660$ pixels with parameters $\log _ { 10 } \left( T_ { \rm vir }\right)=4.101$ and $\log _ { 10 } \left( \zeta\right)=1.389$. (Second top to bottom) the respective responses with non-linear activation function in the arbitrary color scale in $31\times 31 \times 328$ pixels (right) to the trained kernels in the first convolution layer in the 3D CNN (left). For visualization purpose, the response maps are resized to match the size of the input images. Here we only show four out of 32 such kernels in the first convolution layer.}
    \label{fig:kernel}
\end{figure*}

\begin{figure*}
	\includegraphics[width=\textwidth]{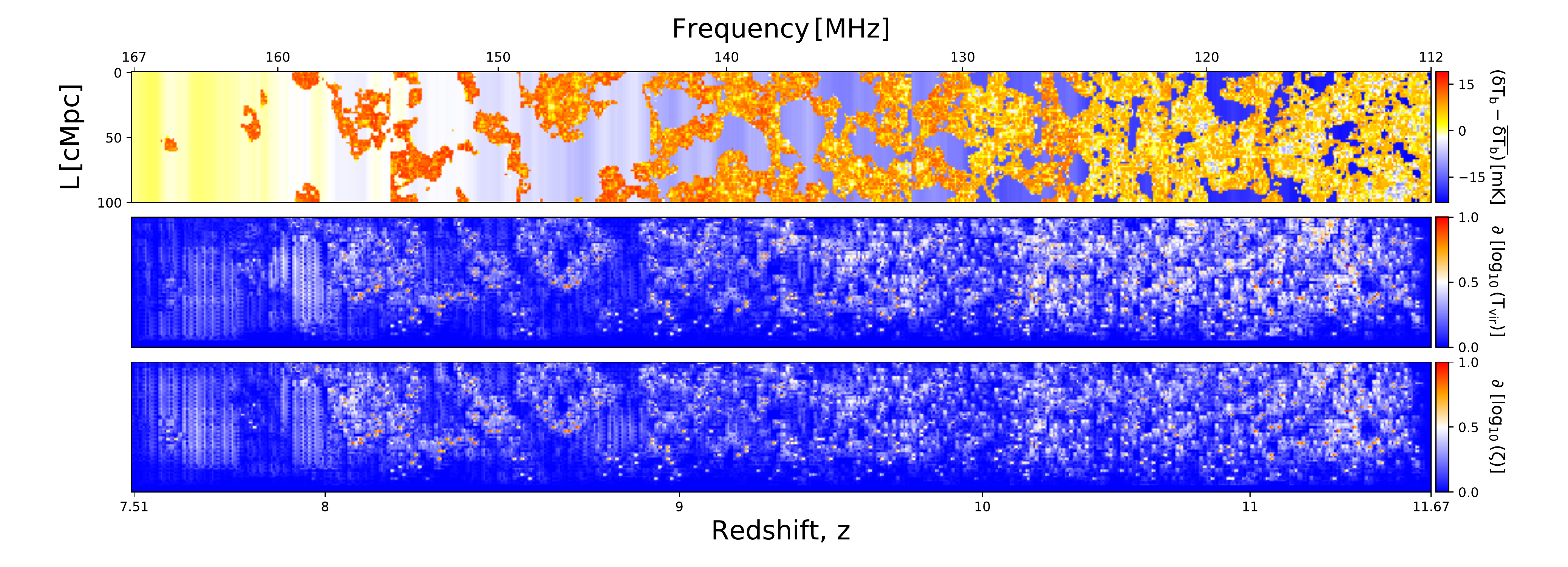}
    \caption{An illustration of saliency maps. (Top) input lightcone 21~cm images with parameters $\log _ { 10 } \left( T_ { \rm vir }\right)=4.101$ and $\log _ { 10 } \left( \zeta\right)=1.389$. (Middle and bottom) the saliency maps of reionization parameters, i.e.\ the logarithmic absolute value of the gradients of the parameter $\log _ { 10 } \left( T_ { \rm vir }\right)$ and $\log _ { 10 } \left( \zeta\right)$, respectively, with respect to the variation of the input image. We rescale the saliency maps to the range $[0, 1]$. } 
    \label{fig:saliency}
\end{figure*}

\begin{table}
    \centering
    \caption{Recovery performance by the 3D CNN and DELFI-3D CNN.}
    \begin{tabular}{cccc}
    \hline \hline
  {}  & {} & 3D CNN & DELFI-3D CNN \\
        \hline
\multirow{2}{*}{$\mathrm{R^2}$\,\tablenotemark{a}} & $ \log_{10}(T_{\rm vir})$ & 0.993 & 0.997\\ 
{} & $\log_{10}(\zeta)$ & 0.983 & 0.992\\
    \hline
 \multirow{2}{*}{$\epsilon$ \,\tablenotemark{b}} & $ T_{\rm vir}$ & $(-0.09, 0.08)$ & $(-0.07, 0.07)$   \\ 
{} & $\zeta$ & $(-0.12, 0.08)$ & $(-0.06, 0.06)$ \\
\hline 
    \end{tabular}
    \flushleft
    \tablenotetext{a}{The coefficient of determination $\mathrm{R^2}$ is computed for the recovered parameter in the logarithmic scale predicted from the networks.}
    \tablenotetext{b}{The fractional error $\epsilon$ refers to that of the deduced parameter in the linear scale within the $68\%$ probability regions in the joint $T_{\rm vir}$-$\zeta$ probability density distribution of their accuracies.}
  \label{tab:perf_comparison}
\end{table}

\section{Results}
\label{sec:results}

\subsection{Parameter Estimation with 3D CNN}
\label{sec:results 3D}

After the 3D CNN is trained with 8000 training samples, it is tested against 1000 test samples. For each test sample, we compare the output reionization parameter set predicted from the 3D CNN with the true parameter value of this sample. The overall recovery performance can be evaluated by the coefficient of determination $\mathrm{R}^2$, defined as  
\begin{equation}
\mathrm { R } ^ { 2 }  = 1 - \frac { \sum \left( y _ { \mathrm { pred } } - y _ { \mathrm { true } } \right) ^ { 2 } } { \sum \left( y _ { \mathrm { true } } - \overline { y } _ { \mathrm { true } } \right) ^ { 2 } }\,,
	\label{eq:R2}
\end{equation}
where $y _ { \mathrm { true } }$ and $y _ { \mathrm { pred } }$ are the true and predicted parameter value for a variable $y$ in a sample, respectively, and the summation for this variable is over all test samples. $\overline { y } _ { \mathrm { true }}$ is the average of true value in all test samples. In general, a score of $\mathrm { R } ^ { \mathrm { 2 } }$ closer to unity indicates a better overall recovery performance of this variable. For our trained 3D CNN, $\mathrm { R } ^ { \mathrm { 2 } } = 0.993 $ and 0.983 for $\log(T_{\rm vir })$ and $\log(\zeta)$, respectively. (\citetalias{Gillet2019} also used the $\mathrm { R } ^ { \mathrm { 2 } }$ value to evaluate the output performance. However, they recover $\zeta$, instead of $\log(\zeta)$.  Our result for $\zeta$ is $\mathrm { R } ^ { \mathrm { 2 } } = 0.953$.)  

We further test the reionization parameter recovery performance by comparing the predicted parameter values from the trained 3D CNN with the true values in Fig.~\ref{fig:comp}. Our results show that the recovery is mostly near the perfect, zero-error, recovery (diagonal line), with reasonably small scatter (which will be quantified by the fractional error below). The side histograms show the distribution  $ p ( u | y _ { \mathrm { pred } } )$ at a given predicted value $y _ { \mathrm { pred }}$, where the residual $u  = y_{\rm pred}  - y _{\rm true}$. In 68\% probability interval, $u \lesssim 0.05$, which means that the fractional error in $T_{\rm vir}$ or $\zeta$ is about $0.05/\lg(e)\approx 12\%$, since the output parameter $y$ is in the logarithmic scale of base $10$. We also note in Fig.~\ref{fig:comp} that near both upper and lower limits of the parameter range, the scatters from the perfect recovery become larger, which means worse recovery accuracy. This is probably due to boundary effects in the parameter space, i.e.\ the networks can not take the data outside the boundaries. This could be avoided in practice by choosing the range of parameter space wide enough around the most likely values indicated from a trial analysis. 

The recovery performance test can be further made by plotting in Fig.~\ref{fig:fractional error} the 2D joint probability density distribution of the fractional errors, $\epsilon = (  y _ { \mathrm { pred } } - y _ { \mathrm { true } }) / y _ { \mathrm { true } }$, where $y$ represents either the parameters directly predicted from the CNN (in the logarithmic scale), $ \log(T_{\rm vir})$ and $\log(\zeta)$, or the deduced parameter (in the linear scale), $ T_{\rm vir}$ and $\zeta$. We employ {\tt GetDist} \citep{Lewis:2019xzd} to make the contour plots throughout this paper. Within the $68\%$ probability regions in the joint probability density distribution, the fractional error with respect to the true value is $0.01$ for $\log _ { 10 } \left( T  _ { \mathrm { vir } } \right)$ and $0.03$ for $\mathrm{log_{10}(\zeta)}$, or  $0.09$ for $T_{\rm vir}$ and $0.12$ for $\zeta$. 

The above tests demonstrate that the 3D CNN can recover the reionization parameters within reasonable accuracy ($\lesssim 12\%$ error) , albeit not perfectly. Between the two parameters, $T_ { \rm vir }$ is recovered with better accuracy than $\zeta$, even if the output weight for $\log(T_ { \rm vir })$ (0.8) is smaller than that for $\log(\zeta)$ (1.0). This implies that the 21~cm images are more sensitive to $T_ { \rm vir }$ than $\zeta$. 
This may be explained in Fig.~\ref{fig:vary_params} by showing the lightcone 21~cm images with varying reionization parameters $\log_{10}(T_ { \rm vir })$ (top panel) and $\log_{10}(\zeta)$ (bottom panel), respectively. For the sake of fair comparison, the step size for both parameters are the same, i.e.\  $\Delta \log_{10}(T_ { \rm vir }) = \Delta \log_{10}(\zeta) = 0.233$. Both decreasing $T_ { \rm vir }$ and increasing $\zeta$ can speed up cosmic reionization. While their impacts on the images are very similar, we find that changing $\log_{10}(T_ { \rm vir })$ affects the lightcone to a slightly larger extent than changing $\log_{10}(\zeta)$, when one carefully compares the morphology of H~{\small II} bubbles. More quantitatively, the mean neutral fraction of hydrogen $\bar{x}_{\rm HI} = 0.496$ at $z=7.51$ (the redshift to the very left of the shown lightcone) for the benchmark model (middle lightcone in both top and bottom panels), but when the parameters are changed to $\log_{10}(T_ { \rm vir }) = 5.466$ ($\log_{10}(\zeta) = 1.233$), the reionization is delayed so that $\bar{x}_{\rm HI} = 0.938$ ($\bar{x}_{\rm HI} = 0.904$) at $z=7.51$, and when the parameters are changed to $\log_{10}(T_ { \rm vir }) = 4.534$ ($\log_{10}(\zeta) = 2.165$), the reionization is accelerated so that $\bar{x}_{\rm HI} = 0.015$ ($\bar{x}_{\rm HI} = 0.018$) at $z=7.51$. These comparisons show that the lightcone 21~cm images are more sensitive to the parameter $T_ { \rm vir }$ than $\zeta$, which explains why the former has better estimation performance with 3D CNN than the latter.

To provide an interpretation of the features the 3D CNNs have learned, we consider two visualization methods as follows. First, we visualize the trained cubic ($5\times 5 \times 5$) kernels for the first layer of convolution, as shown in Fig.~\ref{fig:kernel}. While patterns in the response maps to the kernels are generally difficult to interpret, some patterns might be related to different shapes of H~{\small II} bubbles. We also find some blank response maps that capture no features of the input images, which might imply that some employed kernels may be unnecessary. We do not show the visualization of kernels in the higher-level convolution layers because it is even harder to interpret the information from there.

Saliency mapping \citep{Simonyan2014DeepIC, springenberg2014striving, zeiler2014visualizing, zhou2016learning, selvaraju2017grad} is another useful visualization approach for deep learning interpretation. In this paper, we calculate the simple gradient based saliency maps, which are basically the gradients $\partial \mathbf{t}/\partial \mathbf{d}$ of output data summaries $\mathbf{t}$ with respect to the input field $\mathbf{d}$. Saliency maps visualize the sensitivity of the outputs to the changes of inputs. In Fig.~\ref{fig:saliency}, we plot the saliency maps of the output reionization parameters, $\mathbf{t}_i = \log _ { 10 } \left( T_ { \rm vir }\right)$ and $\log _ { 10 } \left( \zeta\right)$, respectively, with respect to the input lightcone 21~cm images. For visualization purpose, we plot the logarithmic absolute value of the gradients which is rescaled to the range $[0,1]$. We find that the saliency maps are highlighted in the regions close to the boundaries of H~{\small II} bubbles, especially at the middle and late stages of reionization. 

In sum, these visualizations suggest that the information the 3D CNNs extract from the input lightcone images is likely the shape and boundaries of H~{\small II} regions where the fields have sharp changes, both in two spatial dimensions and along one frequency dimension of the input map.

\begin{figure*}
	\centering
	\includegraphics[width=0.8\textwidth]{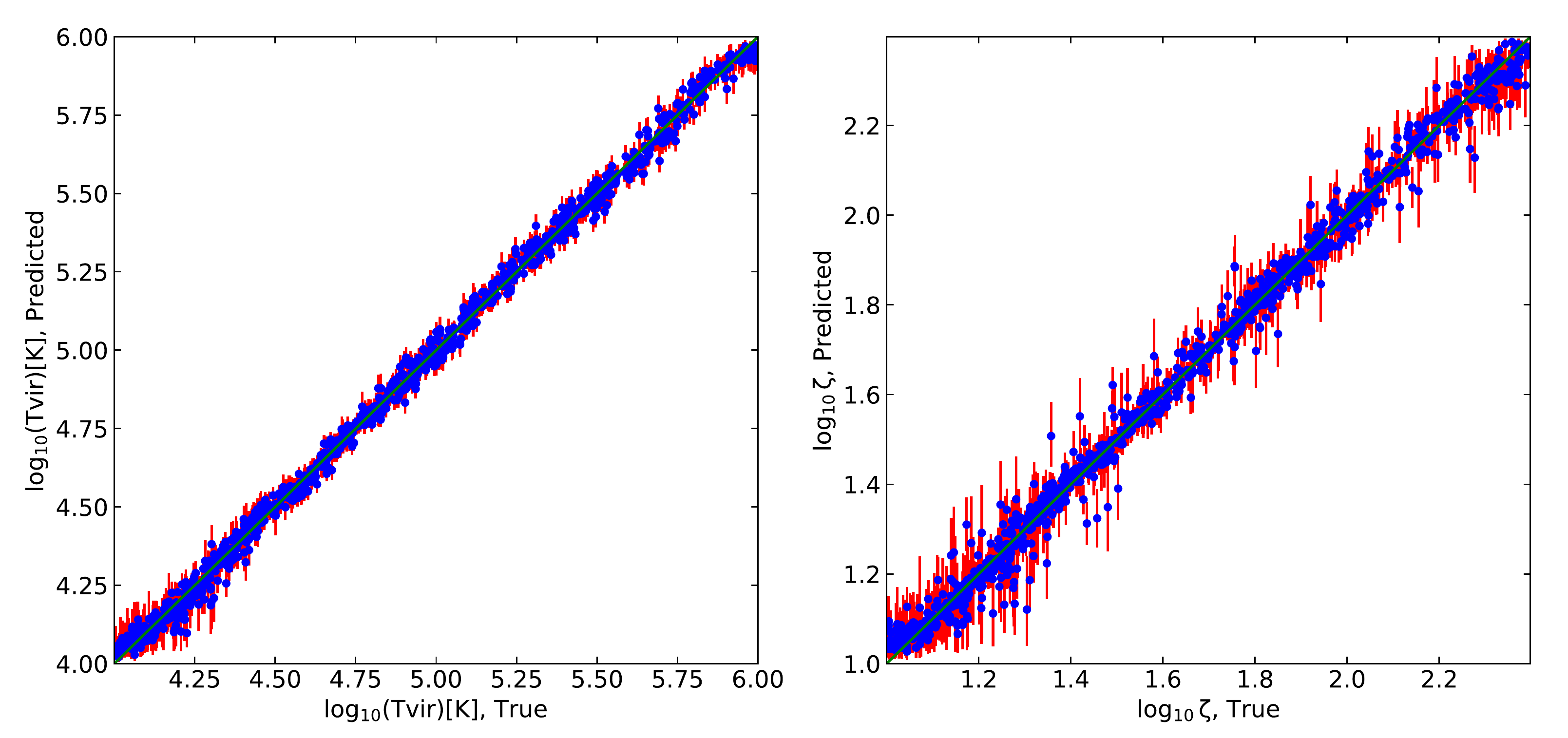}
    \caption{Reionization parameter recovery performance with DELFI-3D CNN using 1000 new test samples. Shown are the predicted value $y_{\rm pred}$ vs the true value $y_{\rm true}$ (blue dots) of each parameter $y = \log(T_{\rm vir})$ (left) and $\log(\zeta)$ (right), respectively, with $1\sigma$ error inferred by the DELFI-3D CNN (red error bars). The green diagonal line indicates the perfect (zero-error) recovery. } 
    \label{fig:comp2}
\end{figure*}

\begin{figure*}
	\includegraphics[width=\textwidth]{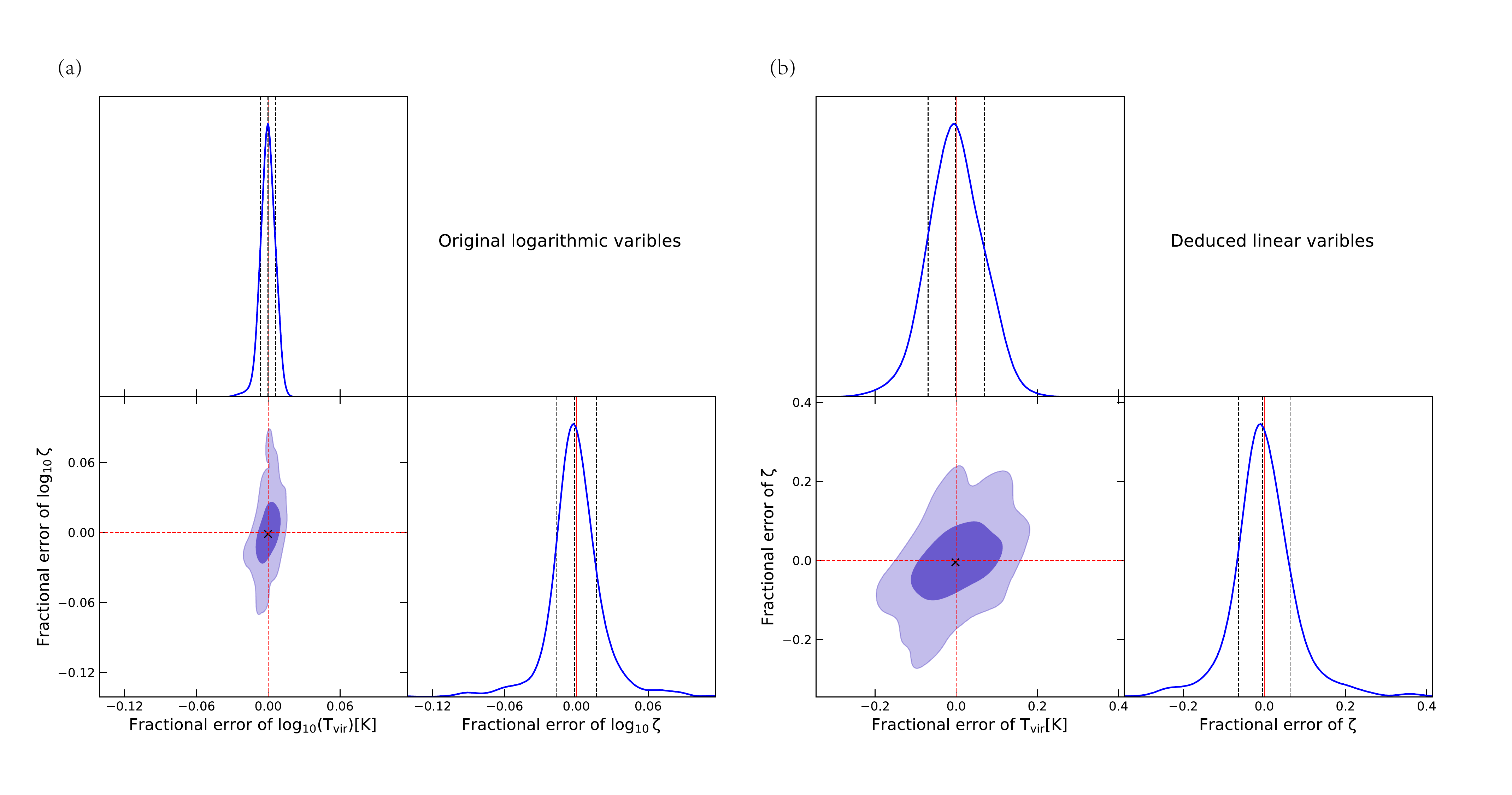}
    \caption{Same as Fig.~\ref{fig:fractional error} but for testing the DELFI-3D CNN with 1000 new test samples. }
    \label{fig:fractional error2}
\end{figure*}

\subsection{Posterior Inference with DELFI-3D CNN}
\label{sec:results delfi}

In order to test on the estimates from the DELFI-3D CNN with the trained NDEs, we employ a new set of 1,000 test samples (different from the test samples for testing the 3D CNN in Section~\ref{sec:results 3D}). From the parameter posterior inferred by  DELFI for each of these test samples, we estimate the posterior median\footnote{An alternative point estimate derived from the posterior would be the posterior mean of the MCMC chains. We find that the $\mathrm { R } ^ { 2 }$ value for the posterior median and that for the posterior mean are nearly identical.} of the MCMC chains and statistical uncertainty. We first test the estimation of parameter medians. In Table~\ref{tab:perf_comparison}, we show that the coefficient of determination is improved from $\mathrm{R^2}=0.993$ ($0.983$) by 3D CNN to $0.997$ ($0.992$) by DELFI-3D CNN, for $\log(T_{\rm vir})$ ($\log\zeta$). In Fig.~\ref{fig:comp2}, we show the predicted parameters with $1\sigma$ errors vs true parameter value, which shows reasonably small scatters from the perfect (zero-error) recovery. Similar to Fig.~\ref{fig:fractional error}, Fig.~\ref{fig:fractional error2} shows the 2D probability density distribution of the fractional errors $\epsilon $ of the parameters estimated by the DELFI-3D CNN with respect to the true values. Within the $68\%$ probability regions in the joint probability density distribution, the fractional error (or the recovery accuracy) is improved from $0.09$ ($0.12$) by 3D CNN to $0.07$ ($0.06$) by DELFI-3D CNN, for $T_{\rm vir}$ ($\zeta$), as shown in Table~\ref{tab:perf_comparison}. These comparisons show that the DELFI-3D CNN can reduce the systematics in the parameter estimation by the 3D CNN. 

\begin{figure*}
	\includegraphics[width=\textwidth]{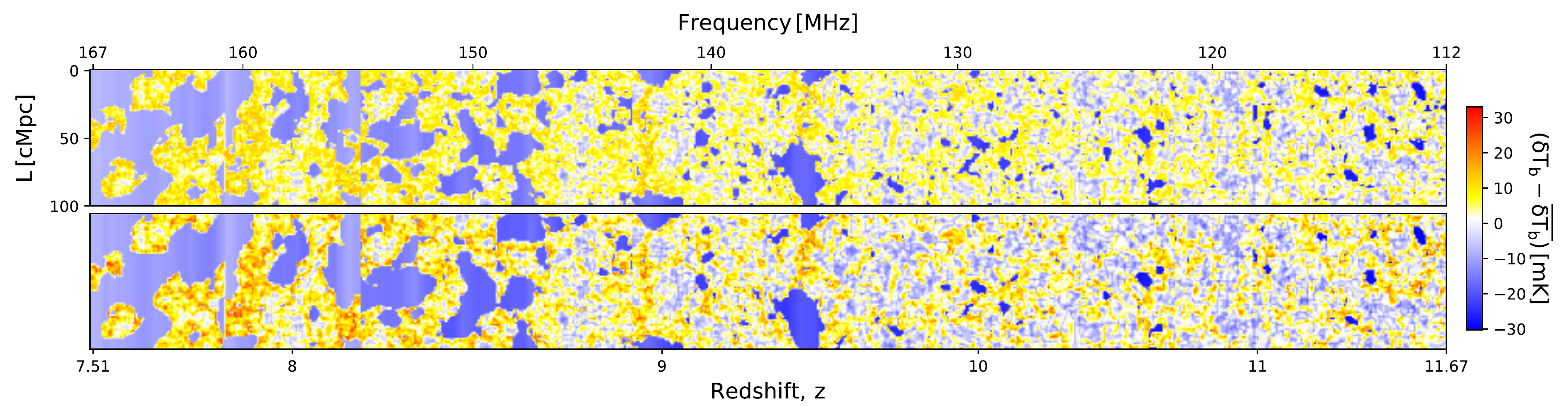}
    \caption{An illustration of the lightcone 21~cm images in two reionization models --- the ``Faint Galaxies Model'' (top), and the ``Bright Galaxies Model'' (bottom), with their true parameter values listed in Table~\ref{tab: bayesian num}.}
    \label{fig:mock}
\end{figure*}

\begin{figure*}
	\centering
	\includegraphics[width=\textwidth]{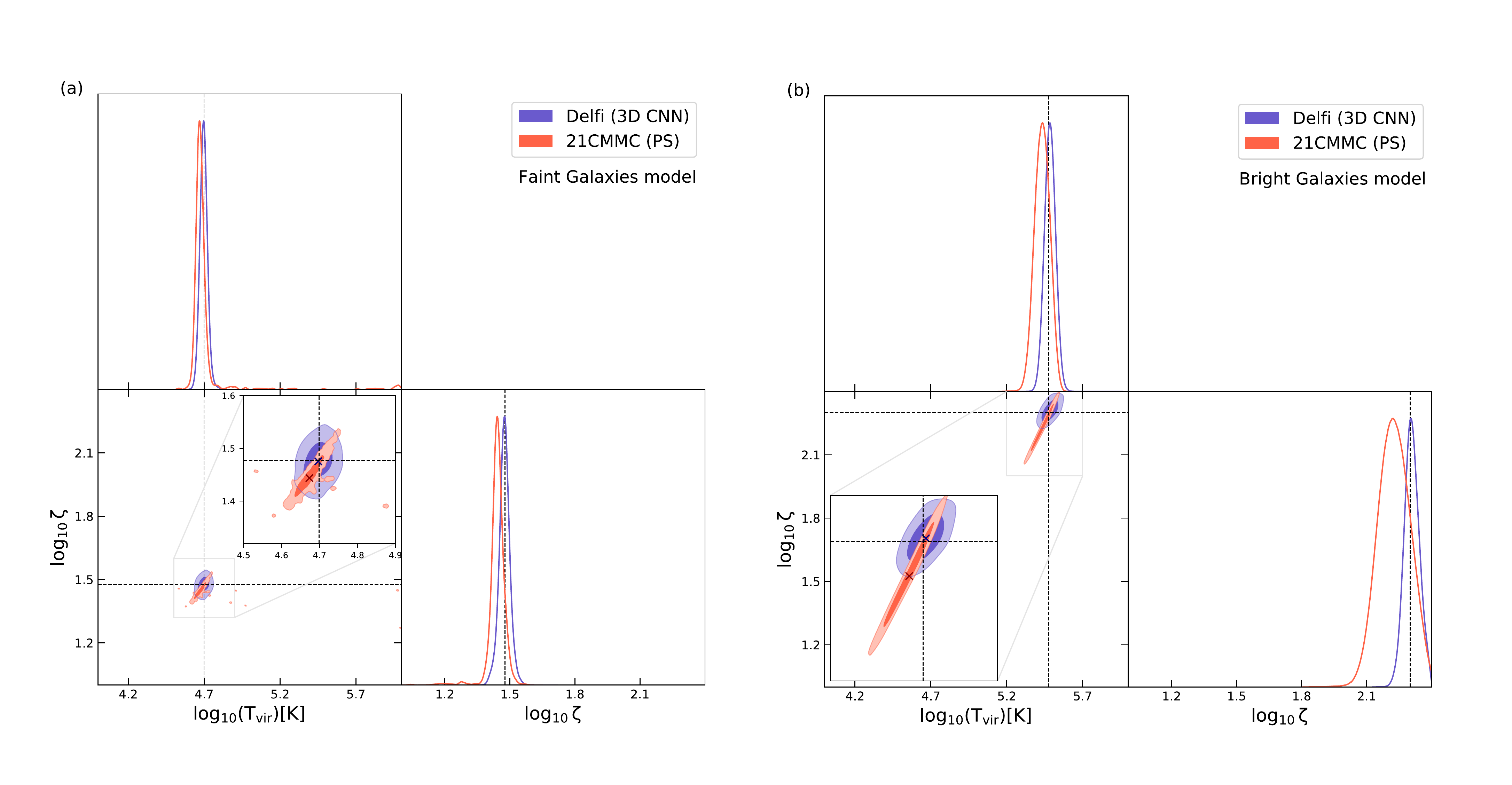}
    \caption{The posteriors estimated by the DELFI-3D CNN (blue) for two mock observations, the ``Faint Galaxies Model'' (left) and the ``Bright Galaxies Model'' (right). We show the median (cross), the $1\sigma$ (dark blue) and $2\sigma$ (light blue) confidence regions. For comparison, we also show the posteriors estimated by the 21~cm power spectrum analysis with MCMC sampling (red). The dashed lines indicate the true parameter values.}
    \label{fig:b1}
\end{figure*}

\begin{table*}
	\centering
	\caption{Bayesian inference with the DELFI-3D CNN and with the {\tt 21CMMC}.}
	\begin{tabular}{ccccccc}
	\hline\hline
			 &\multicolumn{3}{c}{Faint Galaxies Model} &\multicolumn{3}{c}{Bright Galaxies Model} \\
			 \cmidrule(l{.75em}l{.75em}r{.75em}){2-4}
			 \cmidrule(l{.75em}l{.75em}r{.75em}){5-7}
	Parameter & True value & DELFI-3D CNN & {\tt 21CMMC} & True value & DELFI-3D CNN &  {\tt 21CMMC} \\
		 \hline
	$\log _ { 10 } \left( T_ { \rm vir }/{\rm K}\right)$ & 4.699 & $4.697^{+0.024}_{-0.024} $& $4.673^{+0.033}_{-0.025}$ & 5.477 & $5.485^{+0.037}_{-0.036}$& $5.435^{+0.050}_{-0.052}$\\
		  \hline
	$\log _ { 10 }(\zeta)$ & 1.477 & $1.475^{+0.023}_{-0.023}$ & $1.444^{+0.026}_{-0.023}$ & 2.301 & $2.307^{+0.036}_{-0.033}$ & $2.226^{+0.077}_{-0.072}$\\
	 \hline \hline
	$T_ { \rm vir }\,(\times10^5 {\rm K})$ & $0.5$ & $0.498^{+0.029}_{-0.027} $& $0.471^{+0.037}_{-0.026}$ & $3$ & $3.056^{+0.269}_{-0.243}$& $2.724^{+0.331}_{-0.310}$\\
		 \hline
	$\zeta$ & $30$ & $29.9^{+1.6}_{-1.5}$ & $27.8^{+1.7}_{-1.5}$ & $200$ & $202.8^{+17.6}_{-14.8}$ & $168.2^{+32.5}_{-25.6}$\\
	\hline
	\end{tabular}
	\label{tab: bayesian num}
\end{table*}

Now we test the Bayesian inference by the DELFI-3D CNN, a property that cannot be provided by the 3D CNN alone. As a demonstration of concept, we consider two representative mock observations, the ``Faint Galaxies Model'' and ``Bright Galaxies Model'', whose definitions are listed as the ``True value'' in Table~\ref{tab: bayesian num}, following \citet{2017MNRAS.472.2651G}. These models are chosen as two examples with extreme parameter values which however are tuned so that the resulting global reionization histories are both consistent with the latest constraints on the CMB electron scattering optical depth \citep{2016A&A...596A.108P}. As a result, their reionization evolutions are similar, as shown in Fig.~\ref{fig:mock}. However, the H~{\small II} bubbles in the Faint Galaxies Model are smaller and more fractal than in the Bright Galaxies Model, because reionization in the former model is powered by more abundant low-mass galaxies yet with smaller ionization efficiency (due to smaller escape fraction of ionizing photons) than in the latter model. In Fig.~\ref{fig:b1}, we show the results of posterior inference for both models, and list the median and $1\sigma$ errors in Table~\ref{tab: bayesian num}. For the former model, the systematic shift  (i.e.\ relative errors of the predicted medians with respect to the true values) and the $1\sigma$ statistical errors are $-0.04\%\pm 0.5\%$ ($-0.1\%\pm 1.6\%$), for $\log _ { 10 } \left( T_ { \rm vir } \right)$ ($\mathrm{log_{10}\zeta}$), respectively, or equivalently the systematic shift and $1\sigma$ statistical errors are $-0.4\% \pm 5.8\%$ ($-0.3\%\pm 5.4\%$), for $T_{\rm vir}$ ($\zeta$), respectively. For the latter model, the systematic shift and $1\sigma$ statistical errors are  $0.1\% \pm 0.7\%$ ($0.3\% \pm 1.6\%$) for $\log _ { 10 } \left( T_ { \rm vir } \right)$ ($\mathrm{log_{10}\zeta}$), or equivalently $1.9\% {+9.0\% \atop -8.1\%}$ ($1.4\% {+8.8\% \atop -7.4\%} $) for $T_{\rm vir}$ ($\zeta$), respectively. For both models, the total errors are $\lesssim 10\%$ for $T_{\rm vir}$ and $\zeta$. Note that although for the recoveries in Fig.~\ref{fig:b1} the medians happen to be very close to the true values, in general, the medians can be scattered as large as represented by the statistical errors.

Now that we have the tool to estimate the Bayesian posteriors, we can directly test whether more information from the 3D 21~cm images is indeed exploited than just from the 21~cm power spectrum. For the 21~cm power spectrum Bayesian inference with the MCMC sampling, we employ the publicly available code {\tt 21CMMC} \citep{2015MNRAS.449.4246G,2017MNRAS.472.2651G,Greig2018}\footnote{https://github.com/BradGreig/21CMMC}. For the setup of {\tt 21CMMC}, we generate the mock power spectra at 10 different redshifts, each estimated from a coeval\footnote{Strictly speaking, power spectra from the {\it lightcone} boxes should be employed. However, the lightcone effect is only non-negligible at large scales \citep{2012MNRAS.424.1877D,2014MNRAS.442.1491D}, so power spectrum analysis from the coeval boxes, which greatly reduces the computational costs, does not significantly change the inference results. Since the 21~cm power spectrum analysis is not the focus of our paper but just serves for comparison, we choose to use coeval boxes for {\tt 21CMMC} herein.}  
box of $100$ comoving Mpc on each side. In order to perform a fair comparison between the results from the DELFI-3D CNN and from the {\tt 21CMMC}, we take the implementations as follows. (i) These 10 coeval boxes altogether cover the same redshift range with the same reionization parameters as in the 3D CNN. (ii) The modes in the full $k$-range corresponding to the box size and cell size are employed in the {\tt 21CMMC} analysis. (iii) The same set of random seeds for the initial conditions at 10 different coeval boxes is employed for the tests between the DELFI-3D CNN and {\tt 21CMMC}, in order to eliminate any impact of initial conditions. To construct the likelihood for the analysis based on the power spectrum we model the sample variance from the mock observation by $P_{\rm sv} = P_{21}(k) / \sqrt{N(k)}$, 
where $P_{21}(k)$ is the 21~cm brightness temperature power spectrum, and $N(k)$ is the number of modes in a $k$-mode spherical shell. We then perform the Bayesian inference with 200 walkers, using the power spectra at 10 different redshifts. For each walker, we choose the ``burn in chain'' number to be 250, and the main chain number to be 3,000. The results of Bayesian inference for the Faint Galaxies Model and Bright Galaxies Model are shown in Fig.~\ref{fig:b1} and listed in Table~\ref{tab: bayesian num}. 
For the former model, the systematic shift and $1\sigma$ statistical error are $-6\% { +7.9\% \atop -5.5\%}$ ($-7\% {+6.1\% \atop -5.4\%}$), for $T_{\rm vir}$ ($\zeta$), respectively. For the latter model, the systematic shift and $1\sigma$ statistical error are $-9\% {+12.2\% \atop -11.4\%}$ ($-16\% {+19.3\% \atop -15.2\%}$) for $T_{\rm vir}$ ($\zeta$), respectively. We have also checked the convergence of the MCMC chains as shown in Appendix~\ref{app:convergence}. In comparison, the DELFI-3D CNN outperforms the {\tt 21CMMC} both in the median estimation and in the statistical errors of Bayesian inference. This implies that the DELFI-3D CNN may take advantage of more, if not all, information in the 3D 21~cm images than just in the power spectrum. We should also note that the computational speed of DELFI for each inference is two orders of magnitude faster than {\tt 21CMMC}, which makes this approach practically more useful in the futuristic 21~cm data analysis.

\begin{figure*}
	\centering
	\includegraphics[width=0.8\textwidth]{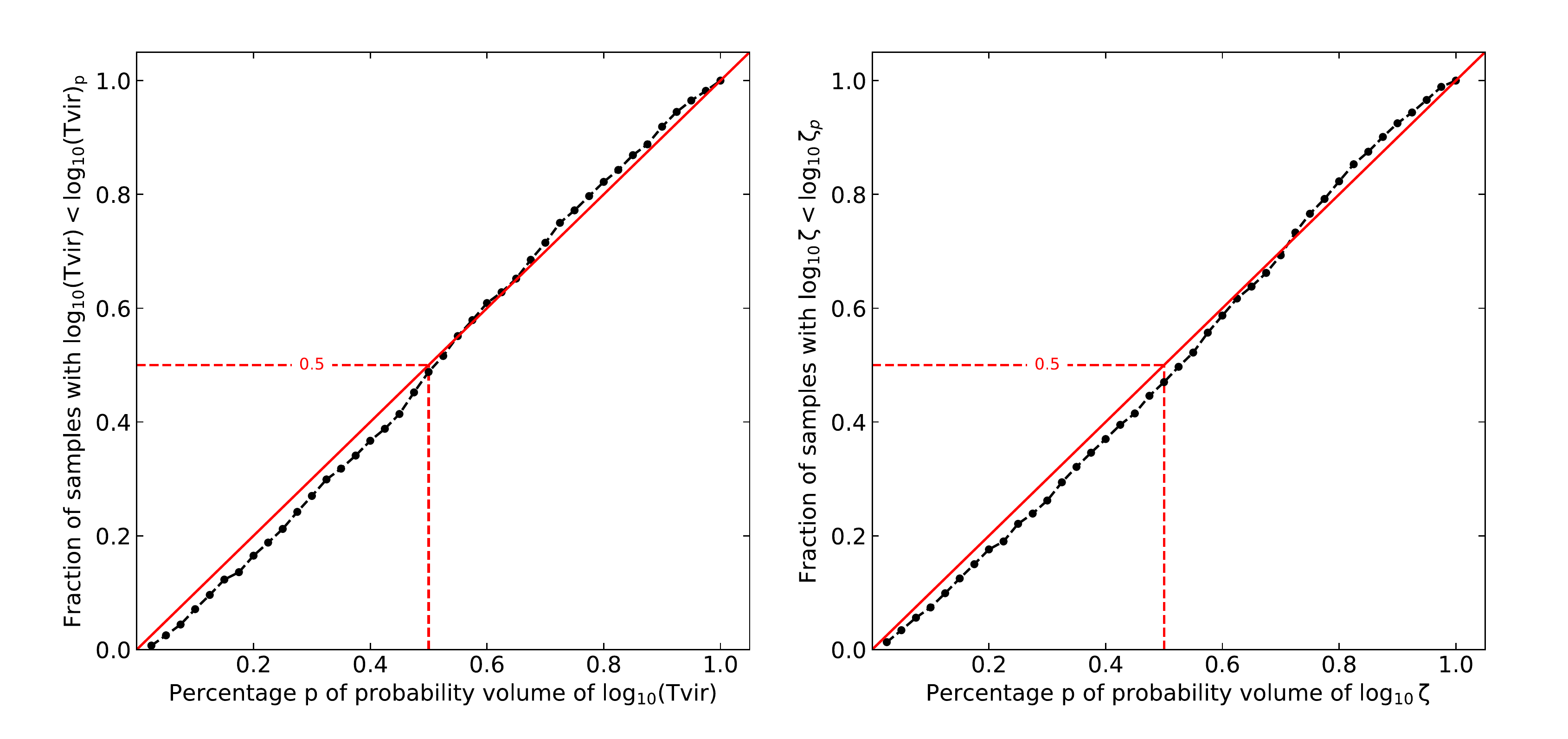}
    \caption{Posterior calibration plot, i.e.\ the empirical percentage $ep$ vs the percentage $p$ of the probability volume for the parameters $\log _ { 10 } \left( T_ { \rm vir }\right)$ (left) and $\log _ { 10 }(\zeta)$ (right). Shown are the actual calibration with 1,000 new test samples (black dots), and the perfect calibration (red solid line).}
    \label{fig:coverage}
\end{figure*}

We further evaluate the accuracy of the posteriors over 1,000 test samples and follow the recent work of \citet{ho2020approximate} to perform the statistical check. Similar evaluation methods could be found in \citet{ramanah2020simulation,levasseur2017uncertainties,wagner2020hierarchical}. In Fig.~\ref{fig:coverage}, we show the posterior calibration plot, which we briefly describe below. For a given percentage $p$ of the probability volume, the fraction of samples with the true parameter values falling into this probability volume of the inferred posterior is defined as {\it empirical percentage} ($ep$), 
\begin{displaymath}
ep=\frac{1}{N} \sum_{i=1}^{N} \mathbf{1}\left[X_{i}<\hat{X}_{p,i}\right]\,,
\end{displaymath}
where $\mathbf{1}[\mathbf{A}]$ is the indicator function which returns 1 if the condition $\mathbf{A}$ is satisfied and 0 otherwise. $N$ is the total number of evaluation samples. Here $X_{i}$ is the true value of the $i$-th sample. $\hat{X}_{p,i}$ is the percentile corresponding to percentage $p$ of the inferred posterior $\hat{x}_{i}$ of the $i$-th  sample, i.e.\ $P(\hat{X}<\hat{X}_{p}|\hat{X} \in \hat{x})=p$. If the the inferred uncertainties are perfectly accurate, then $ep = p$. For univariate posteriors, if the condition $ep>p$ ($ep<p$) always holds, then the inference biases towards larger (smaller) value. If the condition $ep>p$ ($ep<p$) holds when $p<0.5$ and the condition $ep<p$ ($ep>p$) holds when $p>0.5$, then the inference underestimates (overestimates) the uncertainties.

Fig.~\ref{fig:coverage} shows that the uncertainty estimations for both parameters are highly accurate, with only small deviations from the perfect calibration. 
For both parameters, the inferred posteriors are slight overestimation of uncertainties. In other words, our uncertainty estimation is on the conservative side. This trend, as also found in \citet{ramanah2020simulation} (see their Fig.~6) and in \citet{ho2020approximate} (see their Section~4), may result from the non-optimal data compression of the trained 3D CNN.

\section{Summary}
\label{sec:conc}

In this paper, we perform a Bayesian inference of the reionization parameters where the likelihood is implicitly defined through forward simulations using DELFI (specifically the {\tt pydelfi} implementation). The information in the 3D 21~cm images is exploited by compressing the 3D image data into informative summaries with a trained 3D CNN.  
While these summaries are interpretable as point estimates of the reionization parameters, they are technically an intermediate step used as input to full posterior inference of DELFI. We show that this method (DELFI-3D CNN) recovers accurate posterior distributions for the reionization parameters, estimating the posterior median $T_ { \rm vir }$ ($\zeta$) with a relative error of $7\%$ ($6\%$) with respect to the true parameter value, in most ($68\%$) of test samples. 

This level of recovery accuracy improves upon the earlier analysis based on 2D CNN \citepalias{Gillet2019}. In their work, the relative error of the parameter $\log _ { 10 } \left( T  _ { \mathrm { vir } } \right)$ is $\lesssim 1\%$, which is comparable to ours, but the relative error of the parameter $\zeta$ is $\sim 10\%$. 
While the total lightcone volume therein is much larger than ours (300 vs 100 cMpc per each transverse side to the LOS), the 2D study only extracted 5 slices along the LOS for each lightcone. In other words, for the same LOS redshift interval, the data volume exploited in our 3D CNN study is $100^2/(5\times 300)\approx 6.7$ times larger than in the previous 2D study. This explains qualitatively why the DELFI-3D CNN recovery outperforms the 2D CNN study.

For the purpose of comparison, we perform an MCMC analysis of the 21~cm power spectrum alone based on the 3D lightcone and using a Gaussian likelihood approximation. This power spectrum analysis with the {\tt 21CMMC} results in less accurate credible parameter regions than inferred by the DELFI-3D CNN, both in terms of the location and shape of the contours.

These show that the DELFI-3D CNN exploits more information in the 3D 21~cm images than a 2D CNN or just the power spectrum analysis by {\tt 21CMMC}. 

The posterior calibration shows that the uncertainties inferred by the DELFI-3D CNN are statistically self-consistent, but we caution that it is likely that the uncertainties are slightly overestimated.

As a demonstration of concept, this paper only considers the ideal case in which the thermal noise and residual foregrounds are neglected. The DELFI framework is flexible for incorporating these realistic effects, as will be done in a followup paper, so this method will be a promising approach for the scientific interpretation of future 21~cm observation data. We also note that, in principle, 3D CNN can be replaced by other compressors that optimize a trade-off between image compression and information extraction. This provides the room for improving the performance of posterior inference, which will be considered in a followup paper.

\section*{Acknowledgements}
This work is supported by National SKA Program of China (grant No.~2020SKA0110401), NSFC (grant No.~11821303), and National Key R\&D Program of China (grant No.~2018YFA0404502). 
BDW acknowledges support from the Simons Foundation.
We thank Yangyao Chen, Rui Huang, Benoit Semelin, Hayato Shimabukuro and Meng Zhou for useful discussions and helps. The deep learning computations and reionization simulations were ran in the R2D2 and Metis GPU workstations and at the Venus cluster at the Tsinghua University, respectively. 

\software{21CMMC \citep{2015MNRAS.449.4246G,2017MNRAS.472.2651G,Greig2018}, 21cmFAST \citep{Mesinger2007,Mesinger2011}, pydelfi \citep{alsing2019fast}, Keras \citep{chollet2015keras}, GetDist \citep{Lewis:2019xzd}, NumPy \citep{harris2020array}, Matplotlib \citep{Hunter:2007}, SciPy \citep{2020SciPy-NMeth}, scikit-learn \citep{Scikit-learn}, Python2 \citep{van1995python}, Python3 \citep{10.5555/1593511}}

%





\appendix

\section{Optimization of 3D CNN}
\label{sec:opt}

Improving the performance of deep learning 
requires significant tuning. While there is no standard process to achieve the absolutely ``best'' performance, we now reproduce some elements of good practice for tuning to achieve an optimized model that have emerged. 
Interested readers are referred to \citet{GoodBengCour16} for the techniques of network optimization.
First, we can improve the performance by preprocessing the data. 
For example, in order to avoid large values in the networks, we can standardize the data to the range (0,1) (normalization) or (-1,1) (rescaling). 
Secondly, once the data is in place, the hyper-parameters are tuned, e.g. initializing network weights, changing the learning rates and experimenting with different mini-batch sizes and training epochs. 

In this section, we explore the optimization of 3D CNN by considering the variations in network weights, parameter ranges, and sampling strategies, and investigate their impacts on the performance of parameter estimation, as listed in Table~\ref{tab:R2}. The fiducial setup is the one adopted in the main part of this paper. Case A setup focuses on the change of the relative weight of two output parameters, i.e.\ increasing (decreasing) the network weight of $\log _ { 10 } \left( T_ { \rm vir }\right)$ from 0.8 in the fiducial setup to 1.0 (0.6) in Case A1 (A2) when fixing the weight of $\log _ { 10 } \left( \zeta\right)$ to be 1.0. The network weight ratio of 1:1 means equal importance for two output parameters during training, while the network weight ratio of 0.6:1 means that fitting to $\log _ { 10 } \left( T_ { \rm vir }\right)$ is assigned less weight than fitting to $\log _ { 10 } \left( \zeta\right)$. Since the 21~cm signal depends on the output parameters with different sensitivities, a proper adjustment of their network weights can optimize the overall performance of parameter estimation. We find that the weight ratio of 0.8:1 (fiducial setup) gives the best performance results in comparison to Case A1 and A2. 

Next, we consider the effect of the parameter range set {\it a priori}. The range of $\zeta$ is shrunk from $10\le\zeta \le250$ in the fiducial setup to $10\le\zeta\le 100$ in Case B, while fixing the range of $T_ { \rm vir }$ as $10^4 \le T_{\rm vir}\le 10^6\,{\rm K}$. Table~\ref{tab:R2} shows that the accuracies of parameter estimation in fiducial and Case B setups are very similar. 

Thirdly, we consider the sampling of $\zeta$ in the logarithmic scale (Case B) and in the linear scale (Case C) with the same range of $\zeta$, while fixing the sampling of $T_ { \rm vir }$ in the logarithmic scale. Case C has the same sampling strategy as in \citetalias{Gillet2019}. Table~\ref{tab:R2} shows that the accuracies of parameter estimation in Case B and Case C setups are very similar. Note that in Case B, $\mathrm{R^2} = 0.970$ for $\log_{10} \zeta $  corresponds to $\mathrm{R^2} = 0.955$ for $\zeta$, which is almost the same result as in Case C.  


We conclude that different parameter ranges or sampling strategies do not significantly affect the overall performance of parameter estimation with the 3D CNN. However, the performance can be optimized by fine-tuning the relative output weights between different parameters. The fiducial setup in this paper results from this optimization process.

\section{Effect of missing the ${\bf k}_\perp =0 $ mode in the interferometer measurements}
\label{app:hypo}

\begin{figure*}
	\includegraphics[width=\textwidth]{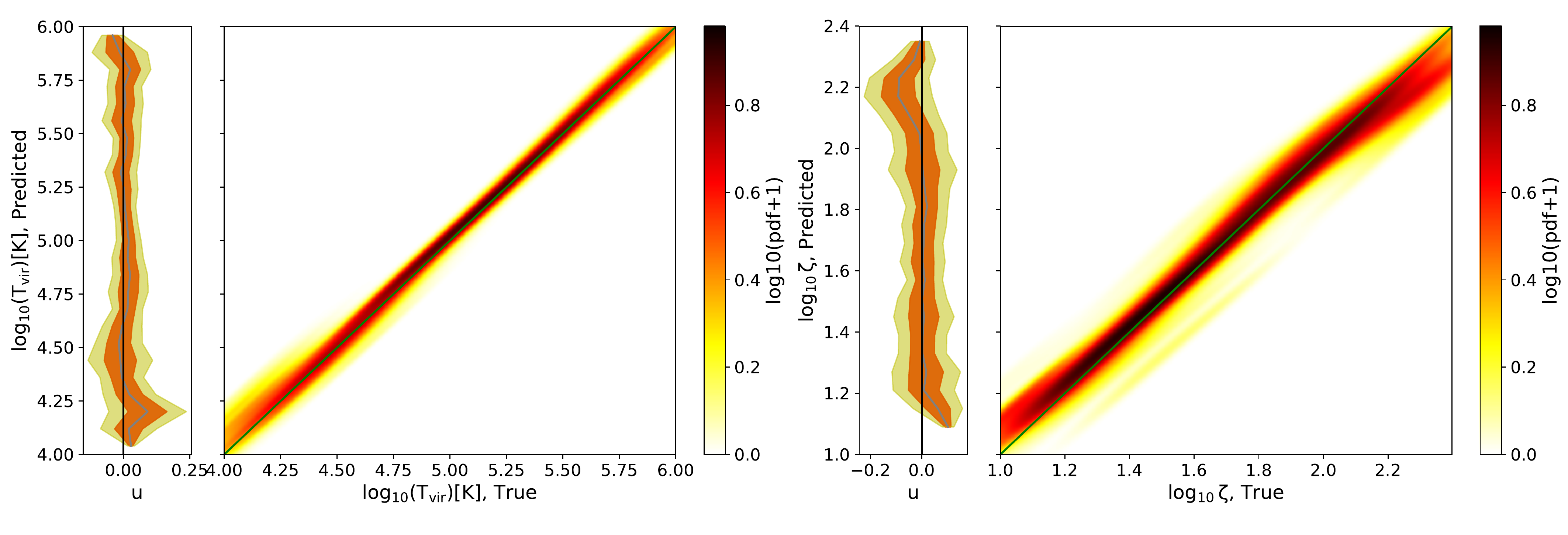}
    \caption{Same as Fig.~\ref{fig:comp} but for the case in which the lightcone images are the 21~cm brightness temperature, without subtraction of the mean of the 2D slice. } 
    \label{fig:comp_hypo}
\end{figure*}

\begin{figure}
	\includegraphics[width=0.5\textwidth]{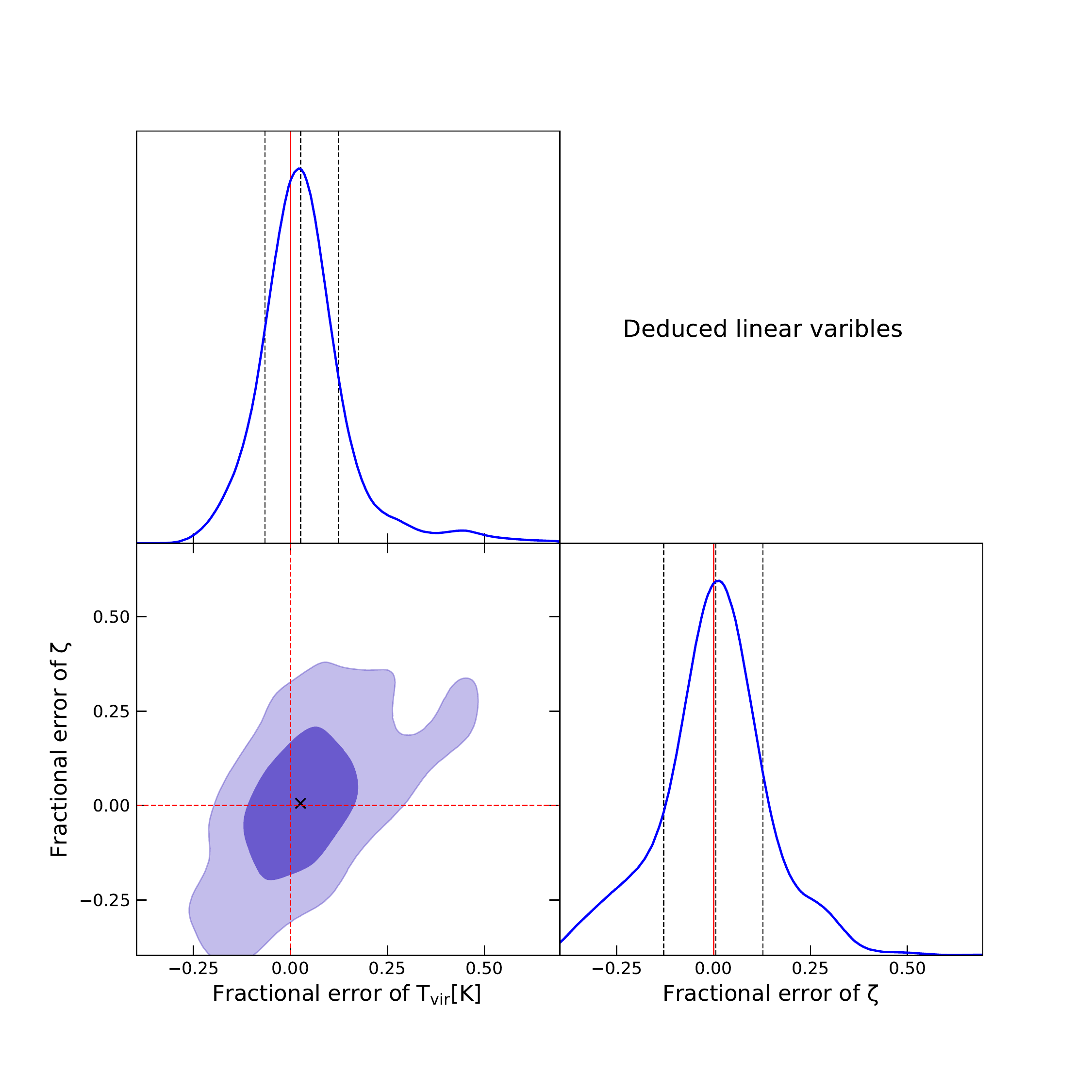}
    \caption{Same as the right panel of Fig.~\ref{fig:fractional error} but for the case in which the lightcone images are the 21~cm brightness temperature, without subtraction of the mean of the 2D slice.  }     \label{fig:fractional error_hypo}
\end{figure}

In our preparation of mock lightcone images, we subtract the mean of the 2D slice from the 21~cm signal, because the interferometer cannot measure the ${\bf k}_\perp =0 $ mode. But in \citetalias{Gillet2019}, this step was skipped. To remove possible systematics due to this effect when comparing the results of \citetalias{Gillet2019} and ours, we perform a test in which the lightcone images are prepared without the subtraction step (see the lower panel of Fig.~\ref{fig:sample}), as in \citetalias{Gillet2019}. We test the reionization parameter recovery performance by comparing the predicted parameter values from the trained 3D CNN with the true values in Fig.~\ref{fig:comp_hypo}, and plotting the 2D joint probability density distribution of the fractional errors for the parameters $T_{\rm vir}$ and $\zeta$ in Fig.~\ref{fig:fractional error_hypo}. 

The $\mathrm{R^2}$ score is 0.992 (0.972) for $\log(T_{\rm vir })$ ($\log\zeta$), respectively.  Within the $68\%$ probability regions in the joint probability density distribution, the fractional error (or the recovery accuracy) is $(-0.12,0.07)$ for $T_{\rm vir}$ and $(-0.13,0.13)$ for $\zeta$, respectively. In comparison with the 3D CNN results shown in Table~\ref{tab:perf_comparison}, the performance of both parameters in this case is similar to (but slightly worse\footnote{Note that the optimization of networks upon the new dataset in the case of not removing the ${\bf k}_\perp =0 $ mode may be different from the fiducial setup. Since this case is not real, we do not tune the networks with greater efforts.} than) that in the fiducial setup of 3D CNN. In other words, not removing the ${\bf k}_\perp =0 $ mode does not affect the performance significantly. 

\section{Formalism of NDEs}
\label{app:ndemath}
\subsection{MDNs}
For the typical Gaussian MDN networks, the density estimators can be defined as:
\begin{equation}
\begin{aligned}
p(\mathbf{t} | \boldsymbol{\theta} ; \mathbf{w})&=\sum_{k=1}^{n_{c}} r_{k}(\boldsymbol{\theta} ; \mathbf{w}) \\
& \times \mathcal{N}\left[\mathbf{t} | \boldsymbol{\mu}_{k}(\boldsymbol{\theta} ; \mathbf{w}), \mathbf{C}_{k} \equiv \mathbf{\Sigma}_{k}(\boldsymbol{\theta} ; \mathbf{w}) \boldsymbol{\Sigma}_{k}^{T}(\boldsymbol{\theta} ; \mathbf{w})\right],
\label{eq:mix}
\end{aligned}
\end{equation}
where $n_{c}$ is the number of Gaussian components. Each component has three properties --- weights of different components $r_{k}(\boldsymbol{\theta}; \mathbf{w})$, means $\boldsymbol{\mu}_{k}(\boldsymbol{\theta} ; \mathbf{w})$, and covariance factors $\mathbf{\Sigma}_{k}(\boldsymbol{\theta} ; \mathbf{w})$. All of these properties are functions of $\boldsymbol{\theta}$ and the network weights $\mathbf{w}$ and different properties require different non-linear activation functions in the output layers of the networks \citepalias{alsing2019fast}. Here the function $\mathcal{N}$ is a multivariate Gaussian distribution.

\subsection{MAFs}

For the vector $\mathbf{t}$ following a condional density $p(\mathbf{t}|\mathbf{\theta})$, we can express $\mathbf{t}$ as a transformation $T$ of $\mathbf{z}$ sampled from a base density $\pi$:
\begin{equation}
\mathbf{t}=T(\mathbf{z}) \quad \text { where } \quad \mathbf{z} \sim \pi(\mathbf{z}|\mathbf{\theta})\,.
\end{equation}

If $T$ is invertible and both $T$ and $T^{-1}$ is differentiable, the conditional density of $p(\mathbf{t}|\mathbf{\theta})$ can be calculated as
\begin{equation}
	\begin{aligned}
		p(\mathbf{t|\mathbf{\theta}})=&\pi\left(\mathbf{z}|\mathbf{\theta}\right)\left|\operatorname{det}\left(\frac{\partial \mathbf{z}}{\partial \mathbf{t}}\right)\right|\\
		=&\pi\left(T^{-1}(\mathbf{t,\theta})|\mathbf{\theta} \right)\left|\operatorname{det}\left(\frac{\partial T^{-1}(\mathbf{t},\mathbf{\theta})}{\partial \mathbf{t}}\right)\right|\,.
	\end{aligned}
\end{equation}

\begin{figure*}
	\includegraphics[width=\textwidth]{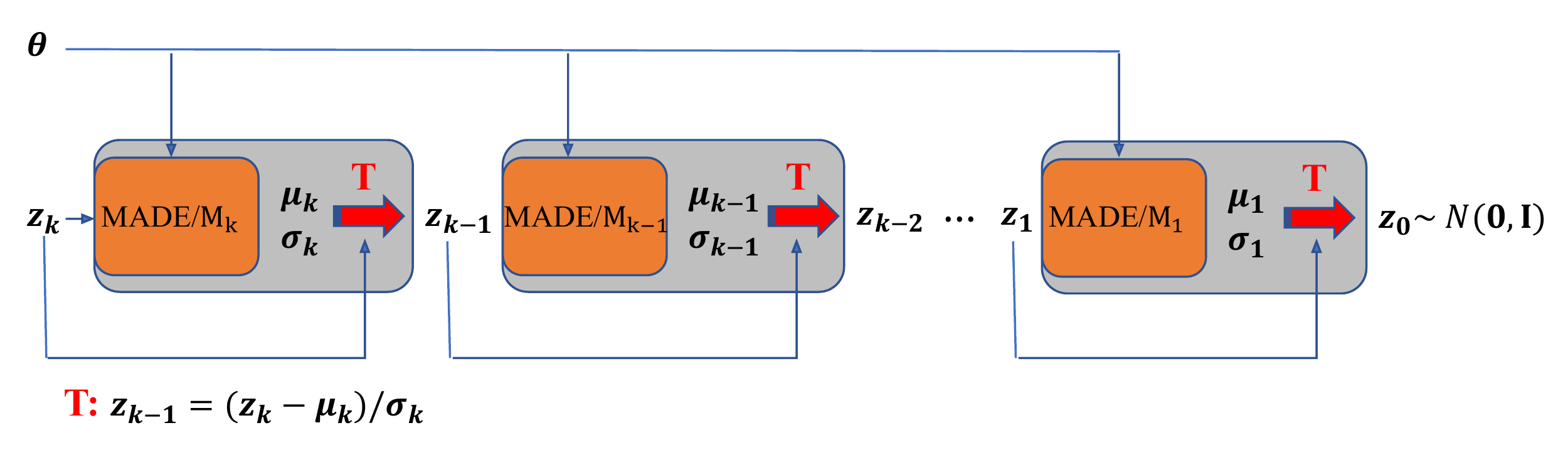}
    \caption{An illustration of the masked autoregressive flows (MAFs). It transforms the data summary $\mathbf{t}$ back to a vector $\mathbf{z_0}$ following a normal distribution. For a single transformation, the MADE takes a vector $\mathbf{z_k}$ and parameter $\mathbf{\theta}$ as the input and outputs the mean $\mathbf{\mu_k}$ and variance $\mathbf{\sigma_k}$ of the autoregressive one-dimensional distributions. Then an affine transformation transforms $\mathbf{z_k}$ into another vector $\mathbf{z_{k-1}}$. Here $\mathbf{z_k}$ refers to the input data summary $\mathbf{t}$ used in this paper.}
    \label{fig:nfs}
\end{figure*}

The masked autoregressive flows (MAFs), as depicted in Fig.~\ref{fig:nfs}, are blocks of single transformations. We first look at a single block of transformation, and simplify the input and output vector as  $\mathbf{t}$ and $\mathbf{z}$, respectively. The MADE with the network weights $\mathbf{w}$ expresses the conditional density in an autoregressive way $p(\mathbf{t}|\mathbf{\theta};\mathbf{w})=\prod_{i=1}^{\operatorname{dim}(\mathbf{t})} p(t_i|t_{1:i-1},\mathbf{\theta};\mathbf{w})$, and ouputs the mean $\mu_i$ and variance $\sigma_i$ of the one-dimensional Gaussian distribution $p(t_i|t_{1:i-1},\mathbf{\theta};\mathbf{w})$:
\begin{equation}
\mu_i=f_{\mu_i}(t_{1:i-1}, \mathbf{\theta};\mathbf{w}),\ \sigma_i=f_{\sigma_i}(t_{1:i-1}, \mathbf{\theta};\mathbf{w})\,.
\end{equation}
Here $f$ represents the MADE. 

An affine transformation can be applied to transform $\mathbf{t}$ to another vector $\mathbf{z}$ with each component
\begin{equation}
	z_{i}=(t_i-\mu_i)/\sigma_i \,.
\end{equation}

In this way, the Jacobin of the transformation is triangular, and the absolute value of determinant can be calculated by
\begin{equation}
	\left|\operatorname{det}\left(\frac{\partial T^{-1}(\mathbf{t},\mathbf{\theta};\mathbf{w})}{\partial \mathbf{t}}\right)\right|=\prod_{i=1}^{\operatorname{dim}(\mathbf{t})}(1/\sigma_{i}(\mathbf{t},\mathbf{\theta};\mathbf{w})) . 
\end{equation}

The composition of single transformations, $T=T_{k} \circ T_{k-1} \circ \ldots \circ T_{1}$, is also invertible and the $T$ and $T^{-1}$ are differentiable. By stacking multiple blocks of transformations, the output of the previous transformation $\mathbf{z_{k-1}}$ is used as the input of the next transformation. The last transformation outputs the vector $\mathbf{z_0}$ that follows the base density which is taken to be a multivariate Gaussian distribution. The final conditional density can be written as
\begin{equation}
\begin{aligned}
	p(\mathbf{t}|\boldsymbol{\theta};\mathbf{w})
	=&\mathcal{N}[\mathbf{z_0}(\mathbf{t}, \boldsymbol{\theta};\mathbf{w}) \mid \mathbf{0}, \mathbf{I}] \\ 
	& \times \prod_{n=1}^{N_{\text {mades }}} \prod_{i=1}^{\operatorname{dim}(\mathbf{t})} (1/\sigma_{i}^{n}(\mathbf{t}, \boldsymbol{\theta};\mathbf{w})),
\end{aligned}
\end{equation}
where the $N_{\text {mades}}$ and $\operatorname{dim}(\mathbf{t})$ are the number of MADEs and the dimension of vector $\mathbf{t}$, respectively.

\subsection{Training of NDEs}
Training NDEs is to fit the estimators $p(\mathbf{t} | \boldsymbol{\theta};\mathbf{w})$ to a target distribution $p^{\star}(\mathbf{t} | \boldsymbol{\theta})$ by minimizing the Kullback-Leibler (KL) divergence between them,
\begin{equation}
D_{\mathrm{KL}}\left(p^{*}|p\right)=\int p^{*}(\mathbf{t} | \boldsymbol{\theta}) \ln \left(\frac{p(\mathbf{t} | \boldsymbol{\theta};\mathbf{w})}{p^{*}(\mathbf{t} | \boldsymbol{\theta})}\right) \mathrm{d} \mathbf{t}\,.
\end{equation}
In practice, we can use a Monte Carlo estimate \citepalias{alsing2019fast}.

\section{Effect of initial conditions}
\label{ICs}

\begin{figure*}
	\includegraphics[width=\textwidth]{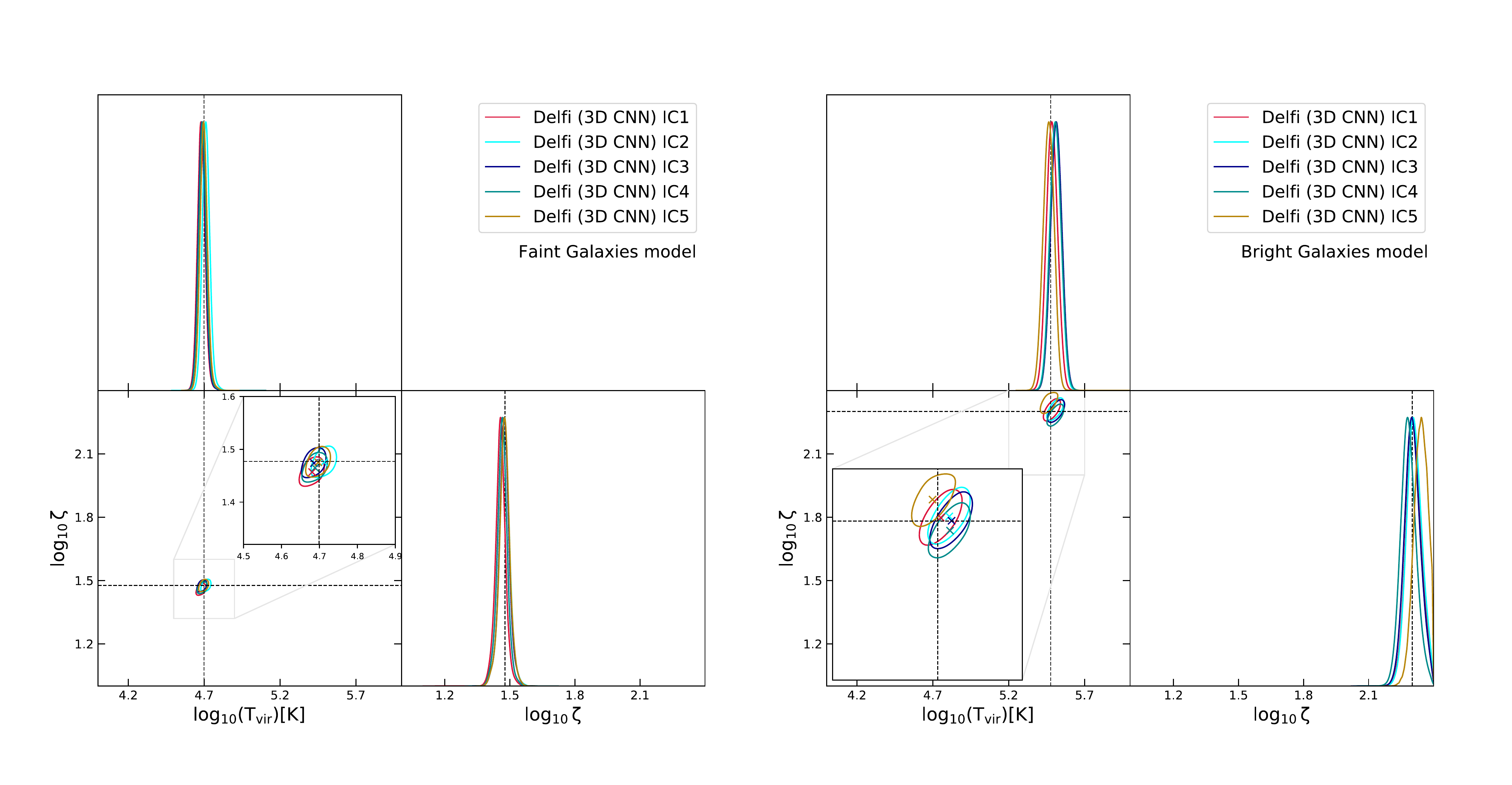}
    \caption{The effect of initial conditions on the posterior inference. Shown are 5 posteriors (in different colors) estimated by the DELFI-3D CNN from the lightcone images with different initial conditions (ICs),  for two mock observations, the ``Faint Galaxies Model'' (left) and the ``Bright Galaxies Model'' (right). We show the median (cross) and $1\sigma$ confidence region (contour). The dashed lines indicate the true parameter values.}
    \label{fig:ics}
\end{figure*}

The 21~cm brightness temperature fields are evolved from the initial Gaussian random density fields, and therefore affected by cosmic variance. In this section, we investigate the impact of different initial conditions of realizations on the parameter recovery with DELFI-3D CNN. In Fig.~\ref{fig:ics}, we plot the posterior inferences from 5 different lightcone images, corresponding to different initial conditions but with the same reionization parameters, for two representative models. We find that, for the Faint Galaxies Model, the $1\sigma$ confidence regions from different initial conditions agree with each other almost completely. For the Bright Galaxies Model, the $1\sigma$ confidence regions show some scatters, but the true values are always within the $1\sigma$ region. We conclude that the posterior inference is mostly robust against the variations in initial conditions due to cosmic variance. 

\section{Stabilization of different NDEs}
\label{app:stack}
\begin{figure}
	\includegraphics[width=0.5\textwidth]{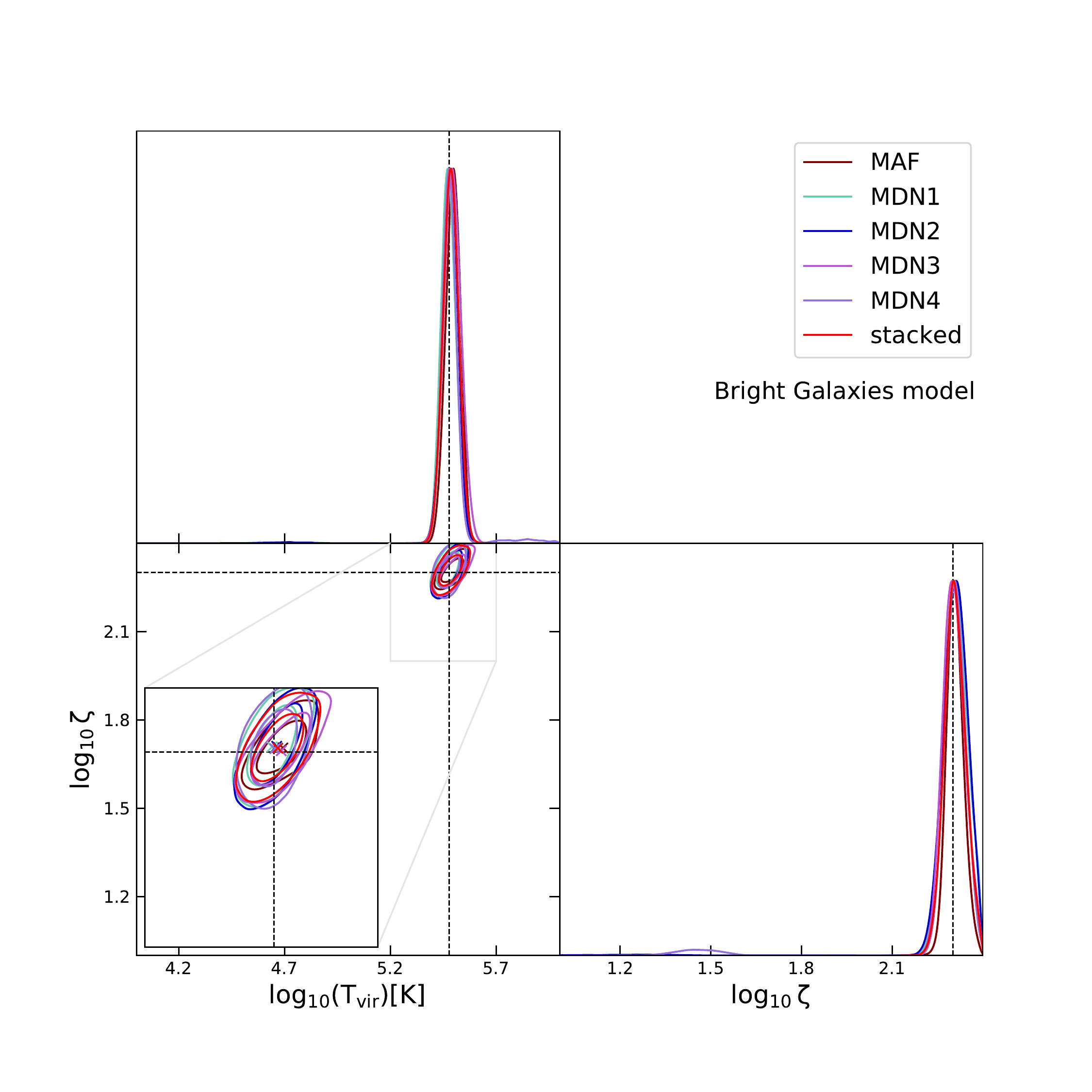}
    \caption{Stabilization of different NDEs. Shown are the posteriors by the DELFI-3D CNN with the individual NDEs (4 MDNs and 1 MAF, with different colors) and with the stacked NDE (red), for the bright galaxies model. We show the median (cross) and $1\sigma$ confidence region (contour). The dashed lines indicate the true parameter values.}
    \label{fig:stacked}
\end{figure}

The default setup of NDE in this paper is a stacked ensemble of individual NDEs, including 4 MDNs and 1 MAF. The resulting posterior, therefore, is stacked from individual posteriors with weights according to the training loss of the corresponding NDEs. In this section, we test on the robustness of posterior inference with different ensembles of NDEs. In Fig.~\ref{fig:stacked}, we show the stacked posterior and the individual posteriors for the bright galaxy model. We find that the $1\sigma$ confidence region inferred with different NDEs agree with each other very well. We conclude that the posterior inference is robust against the variations in the NDEs. 

\section{Convergence of {\tt 21CMMC}}
\label{app:convergence}
\begin{figure*}
	\includegraphics[width=\textwidth]{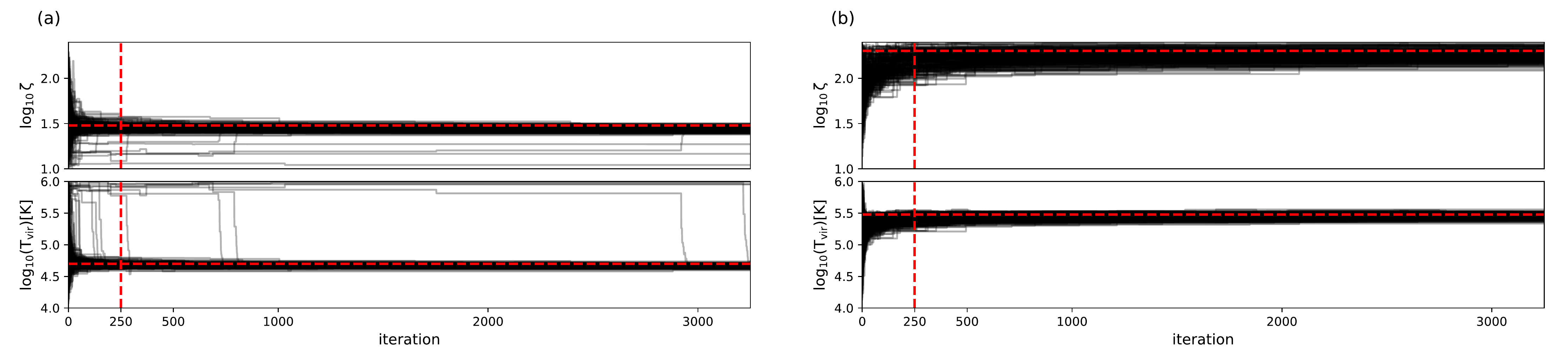}
    \caption{MCMC convergence test for {\tt 21CMMC} for two mock observations --- the ``Faint Galaxies Model'' (left) and the ``Bright Galaxies Model'' (right). For a given mock observation, we use 200 walkers, each with independent trajectory (black lines). We discard the first 250 iterations --- the so-called ``burn in'' of MCMC chains as indicated by the vertical red dashed lines, and only use the following 3000 iterations to generate the {\tt 21CMMC} results. The true parameter values are shown by the horizontal red dashed lines.}
    \label{fig:MCMC}
\end{figure*}

In Fig.~\ref{fig:MCMC}, we make a convergence test for the {\tt 21CMMC} run, by showing the trajectories of 200 walkers for each mock observation. The trajectories represent the updating process of the parameters used to explore the likelihood at the two mock observations. For each walker, we discard the initial 250 iterations (``burn in'') and adopt the subsequent 3000 iterations to generate the {\tt 21CMMC} results. For both models, the trajectories reach the convergence within the burn in iterations, which means that the {\tt 21CMMC} results are converged.

\begin{table*}
	\centering
	\caption{Variations of network setups in the 3D CNN and their results of reionization parameter estimation.}
	\begin{tabular}{cccccccc} 
		\hline\hline
		  Model &  &  &  Fiducial & Case A1 & Case A2 & Case B & Case C \\ 
		  \hline 
		  \multirow{3}{*}{Setup} & & Weight for $\log _ { 10 } \left( T_ { \rm vir }\right)$ \tablenotemark{a} & $0.8$ & $1.0$ & $0.6$ & $0.8$ & $0.8$ \\
		  {} & & Parameter Range for $\zeta$ \tablenotemark{b}  & $1\le \log _ { 10 }\zeta \le 2.398$ & $1\le \log _ { 10 }\zeta \le 2.398$ & $1\le \log _ { 10 }\zeta \le 2.398$ & $1\le \log _ { 10 }\zeta \le 2$ & $10\le \zeta \le 100$ \\
		  {} &  & Output Type for $\zeta$ \tablenotemark{c}   & logarithmic & logarithmic & logarithmic & logarithmic & linear \\
		  \hline
		  \multirow{4}{*}{Results} & \multirow{2}{*}{$\mathrm{R^2}$\,\tablenotemark{d}} & $ \log_{10}(T_{\rm vir})$ & $0.993$ &  $0.991$ &  $0.989$ &  $0.995$ &  $0.991$ \\
		  {} & {} & $ \log_{10}(\zeta)$ & $0.983$ & $0.970$ & $0.977$ & $0.970$ & $0.951$ \\
		  {} & \multirow{2}{*}{$\epsilon$ \,\tablenotemark{e}} & $T_{\rm vir}$ & $(-0.09,0.08)$ & $(-0.12,0.11)$ & $(-0.14,0.10)$ & $(-0.08,0.09)$ & $(-0.15,0.08)$ \\
		  {} & {} & $\zeta$ & $(-0.12,0.08)$ & $(-0.12,0.14)$ & $(-0.11,0.11)$ & $(-0.09,0.08)$  & $(-0.10,0.08)$ \\
		  \hline
        \end{tabular}
    \flushleft
    \tablenotetext{a}{The network weight for $\log _ { 10 }\zeta$ (or $\zeta$ in Case C) is fixed as $1.0$.}
    \tablenotetext{b}{The parameter range for $T_ { \rm vir }$ is fixed as $4 \le \log_{10}(T_{{\rm vir}}/{\rm K}) \le 6$.}
    \tablenotetext{c}{``Logarithmic'' (``linear'') for $\zeta$ refers to the output of $\log _ { 10 }\zeta$ ($\zeta$), respectively. The output for $T_ { \rm vir }$ is always logarithmic, i.e. $\log _ { 10 } \left( T_ { \rm vir }\right)$.}
    \tablenotetext{d}{The coefficient of determination $\mathrm{R^2}$ is computed for the recovered parameter in the logarithmic scale predicted from the networks, except for Case C where $\mathrm{R^2}(\zeta)$ is for $\zeta$ in the linear scale.}
    \tablenotetext{e}{The fractional error $\epsilon$ refers to that of the deduced parameter in the linear scale within the $68\%$ probability regions in the joint $T_{\rm vir}$-$\zeta$ probability density distribution of their accuracies.}
  \label{tab:R2}
\end{table*}

\bibliographystyle{aasjournal}
\bibliography{Ref3}{}



\end{document}